\documentclass[prb,twocolumn,groupedaddress]{revtex4}

\usepackage{amsmath}
\usepackage{amsfonts}
\usepackage{amssymb,mathrsfs}
\usepackage{graphicx}

\begin{document}

\title{Majorana fermions in the nonuniform Ising-Kitaev chain: exact solution}

\author{B.N. Narozhny}

\affiliation{Institut f\"ur Theorie der Kondensierten Materie,
  Karlsruher Institut f\"ur Technologie, 76128 Karlsruhe, Germany}
\affiliation{National Research Nuclear University MEPhI (Moscow
  Engineering Physics Institute), Kashirskoe shosse 31, 115409 Moscow,
  Russia}

\date{\today}

\begin{abstract}

 A quantum computer based on Majorana qubits would contain a large
 number of zero-energy Majorana states. This system can be modelled as
 a connected network of the Ising-Kitaev chains alternating the
 ``trivial'' and ``topological'' regions, with the zero-energy
 Majorana fermions localized at their interfaces. The low-energy
 sector of the theory describing such a network can be formulated in
 terms of leading-order couplings between the Majorana zero modes. I
 consider a minimal model exhibiting effective couplings between four
 Majorana zero modes -- the nonuniform Ising-Kitaev chain, containing
 two ``topological'' regions separated by a ``trivial''
 region. Solving the model exactly, I show that for generic values of
 the model parameters the four zero modes are localized at the four
 interface points of the chain. In the special case where additional
 inversion symmetry is present, the Majorana zero modes are
 ``delocalized'' between two interface points. In both cases, the
 low-energy sector of the theory can be formulated in terms of the
 localized Majorana fermions, but the couplings between some of them
 are independent of their respective separations: the exact solution
 does not support the ``nearest-neighbor'' form of the effective
 low-energy Hamiltonian.

\end{abstract}

\maketitle

\noindent 
Physicists have been fascinated with Majorana fermions ever since
their discovery \cite{Majorana,Wilczek} in 1937, when Ettore Majorana
found a completely real (i.e. not containing complex coefficients)
representation of the Dirac equation. The solutions of the Majorana
equation describe neutral fermions -- particles that obey the Fermi
statistics, but at the same time are their own antiparticles. Whether
they exist in nature as elementary particles is still an open
question. It has been hypothesized that neutrinos might be Majorana
fermions. This hypothesis could be experimentally confirmed by
observation of an elusive process known as the neutrinoless double
beta decay \cite{Avignone}, which is the focus of considerable
experimental efforts.

Recently, it has become possible to imitate the ideas of the
relativistic field theory in solids. Following the success of graphene
research, further novel materials have been identified as Dirac
\cite{Young} and Weyl \cite{Hasan,Yan} semimetals. These materials
exhibit a finite number of band crossings at the Fermi level (the
so-called Dirac and Weyl points) \cite{Balents,Bernevig}. To a good
approximation, low-energy excitations near these points are
characterized by the (quasi)-relativistic spectrum allowing one to
observe phenomena previously belonging to the realm of high energy
physics, such as the Bell-Adler-Jackiw chiral anomaly
\cite{Adler,Bell,Nielsen,Zhang}.

\begin{figure}[t]
\centerline{\includegraphics[width=0.95\columnwidth]{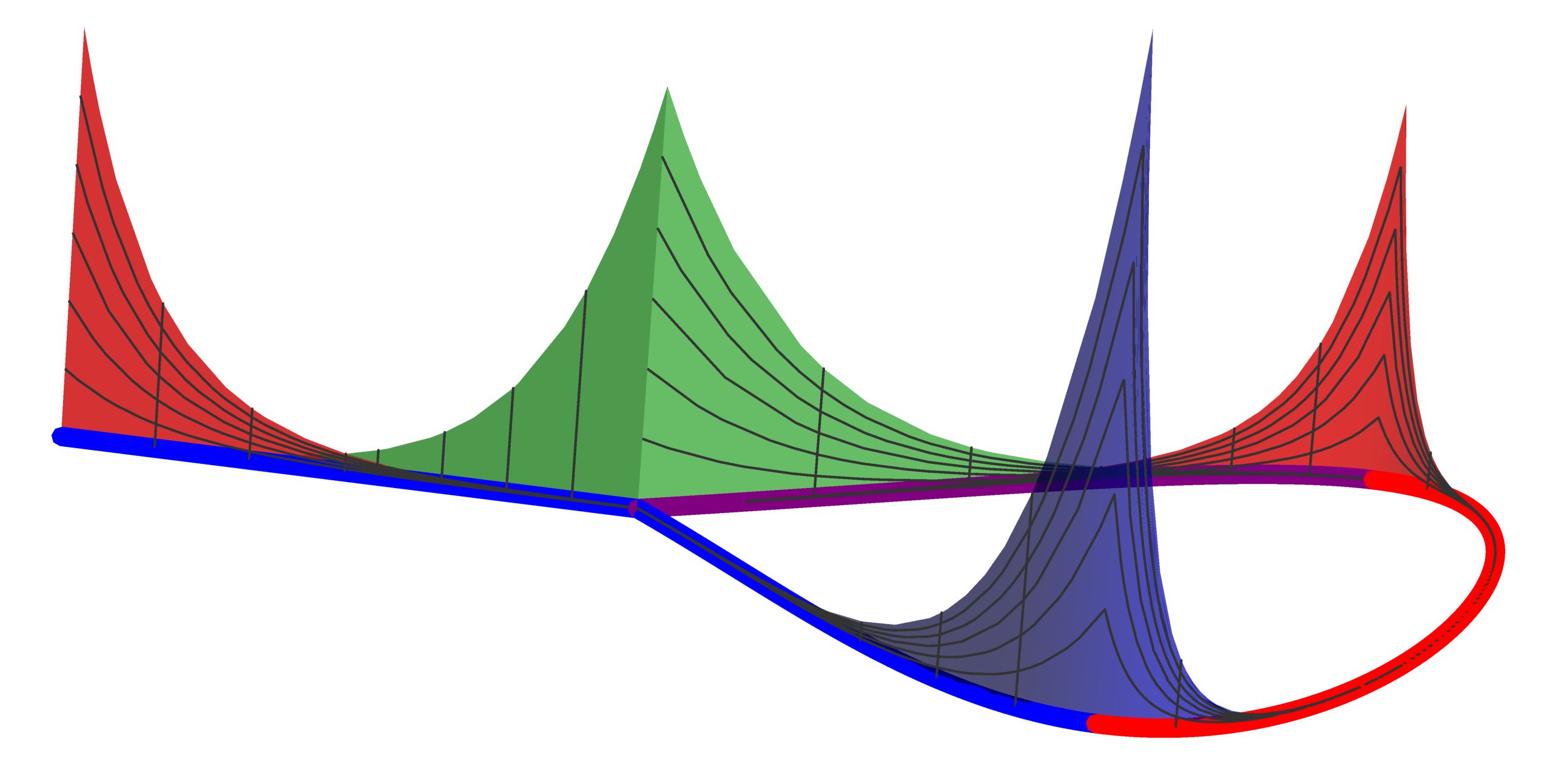}}
\caption{A three-dimensional illustration of the Majorana bound states
  in the Kitaev model with a T-junction. The chain contains two
  ``topological'' regions (the blue line, with $N_1$ sites, and the
  purple line, with $N_2$ sites) and one ``trivial'' region (the red
  line, with $M$ sites). The T-junction is located at the site
  $N_0$. The peaks represent the absolute values of the real-space
  amplitudes of the Majorana zero modes calculated for
  ${N_1\!=\!M\!=\!20}$, ${N_2\!=\!10}$, and ${N_0\!=\!11}$. The model
  exhibits four Majorana zero modes (corresponding to the nearly
  four-fold degeneracy of the ground state). Two of them are localized
  at a single interface point each: the dark blue at the site $N_1$
  and the green at the T-junction. Note, that this amplitude is spread
  over only two (out of three) branches at the junction. The remaining
  two zero modes are delocalized between two interface points, the
  sites $1$ and ${N_1\!+\!M}$. One of them is illustrated by the red
  peaks.}
\label{fig:ab_T_3D}
\end{figure}

At the same time, signatures of Majorana fermions were found in
nanowires with proximity-induced superconductivity
\cite{Mourik,Rokhinson,Heiblum,Deng,Marcus,DeFranceschi,Albrecht,Marcus16}. While
the physics of such systems is rather complex, the effective
low-energy Hamiltonian describing the nanowire is essentially that of
the one-dimensional (1D) {\it p}-wave superconductor, i.e. the
continuous limit of the Kitaev model
\cite{Kitaev,Kane,Lutchyn,Oreg}. The 1D lattice model proposed by
Kitaev \cite{Kitaev} exhibits a quantum phase transition between two
gapful (massive) phases, known as the ``trivial'' and ``topological''.
The ``trivial'' phase is characterized by a single non-degenerate
ground state, while the ``topological'' phase possesses a ground state
that is nearly doubly degenerate: for any finite-size, open chain the
difference between the energies of the lowest-lying excited state and
the ground state is exponentially small, ${\propto\exp(-L/\ell_0)}$,
in the length of the chain $L$ (here $\ell_0$ is a certain correlation
length defined below). In the thermodynamic limit, the energy
difference vanishes and the ground state becomes truly
degenerate. This is a manifestation of a well known theorem in
statistical physics \cite{Barber}: spontaneous symmetry breaking and
the corresponding vacuum degeneracy may only occur in the
thermodynamic limit. The lowest-lying excitation in the
``topological'' phase of the Kitaev model is a fermion with a wave
function that is nonzero (with exponential accuracy) only near the two
edges of the chain \cite{Kitaev}. This fermion can be described in
terms of two Majorana fermions, one at each edge
\cite{Kitaev,Beenakker}. It is these objects, known as the ``Majorana
bound states'' \cite{Mourik,Oreg,Marcus16} or ``Majorana zero modes''
\cite{Rokhinson,Marcus,Albrecht}, that have been observed.

Arguably the main driving force behind the pursuit of Majorana zero
modes in solids is the possibility of applications to quantum
computing \cite{Kitaev,Alicea,Kitaev08}. The basic building block of a
quantum computer, the qubit, can be realized as a coupled system of
four spatially separated zero-energy Majorana states
\cite{Beenakker,Nayak,Nori15}. It is expected that a Majorana qubit
would have a rather long coherence time due to its topological nature
\cite{Nayak,Stern}. Quantum computer can then be envisioned as a
connected network of such qubits. Certain logical operations in such a
computer can also be performed topologically by means of braiding (or
adiabatic interchange) of Majorana fermions \cite{Nayak,Halperin}.

Alternatively, one can search for Majorana fermions in manifestly
discreet systems \cite{Mooij,Nori,Oreg13,Nori16}. For instance, one
may engineer the Majorana bound states using Josephson qubits
\cite{Ioffe,Nori11} to build an artificial spin chain
\cite{Mooij,Nori} that is designed to be an experimental realization
of the 1D quantum Ising model
\cite{Suzuki,Shura,Tsvelik,Lieb,Pfeuty}. The quantum Ising chain with
open boundary conditions is formally equivalent to the Kitaev chain
\cite{Kitaev,Kitaev08,Nori,Lieb,Pfeuty} (note, that the two models do
not enjoy the same level of the topological protection
\cite{Kitaev,Greiter}). This equivalency is based on the Jordan-Wigner
transformation \cite{JW} that is commonly used in 1D theories to
express the spin-1/2 operators in terms of creation and annihilation
operators of spinless fermions \cite{Shura}. In fact, the original
solution \cite{Pfeuty} of the 1D quantum Ising model was based on the
consequent application of the Jordan-Wigner transformation and the
Bogolyubov transformation \cite{Bog}, mapping the model onto a system
of free fermions \cite{Lieb}. The simplicity of the resulting physical
picture may be deceptive, since both the Jordan-Wigner and Bogolyubov
transformations are nonlocal \cite{Greiter}. Although the original
Hamiltonian contains only nearest-neighbor couplings, the model may
develop long-range correlations. In fact, the ground state of the
open-ended chain is characterized by the ``end-to-end'' correlation
function \cite{Pfeuty} that vanishes in the ``trivial'' phase (in the
thermodynamic limit), but remains finite in the ``topological''
phase. This result can be interpreted in terms of a nonlocal fermion
operator that is a linear combination of the Jordan-Wigner fermions at
both ends of the chain.  The lowest excited state of the open-ended
chain in the ``topological'' phase (i.e. the state that is nearly
degenerate with the ground state) possesses a similar structure. The
wave function of this state decays exponentially away from the chain
ends and hence can be represented as a linear combination of the two
states localized at either end of the chain. The existence of such
edge states has been known for a very long time \cite{Lieb}, but they
were not interpreted in terms of Majorana fermions and related to the
quantum information theory before the work of Kitaev \cite{Kitaev}.

A quantum computer based on Majorana qubits would contain a large
number of Majorana zero modes. Whether the device will be built using
the nanowires \cite{Kane,Lutchyn,Oreg} or the artificial spin chains
\cite{Mooij,Nori}, one can envision the effective model of the system
as a connected network of the Ising-Kitaev chains alternating the
``trivial'' and ``topological'' regions, with the zero-energy Majorana
fermions localized at their interfaces \cite{Alicea}. The low-energy
sector of such a theory can be formulated in terms of leading-order
couplings between the Majorana zero modes
\cite{Alicea,Shnirman,Nori15,Wang}. These couplings are often chosen
based on physical intuition. Given the nonlocal relation between the
Majorana zero modes and the Kitaev (or Jordan-Wigner) fermions, it is
desirable to test that intuition against a rigorous solution of a
representative microscopic model. This is the principle goal of the
present work.

In this paper I consider a minimal model exhibiting effective
couplings between Majorana zero modes -- the nonuniform Ising-Kitaev
chain, containing two ``topological'' regions separated by a
``trivial'' region. Based on the common intuition, one would expect
that this model possesses four Majorana zero modes, each localized at
one of the four interface points of the chain \cite{Alicea,Shnirman}
(i.e. the two chain ends and two edges of the ``trivial'' region). I
present the exact solution of the model and identify the region of
model parameters where the above expectation is indeed
fulfilled. However, the exact solution also exhibits situations where
the intuitive expectation is {\it not} fulfilled. In particular,
inversion symmetry (in the case where the two ``topological'' regions
are identical) leads to ``delocalization'' of the Majorana zero modes
between two interface points. While one can use a basis rotation to
express the low-energy sector of the theory in terms of four localized
Majorana operators, the corresponding states will no longer be the
eigenstates of the model. The low-energy Hamiltonian will then contain
effective couplings between some of these modes that are independent
of their respective separations. I also demonstrate that the symmetric
case in not the only situation exhibiting the ``delocalization'' of
the Majorana bound states. As an example, I show that the
``delocalization'' may also occur in the variant of the model, where
one of the chain ends is coupled to one of the intermediate sites
forming a T-junction (or a Y-junction), see Fig.~\ref{fig:ab_T_3D}.

\section{The nonuniform Ising-Kitaev chain}

The open-ended, nonuniform quantum Ising chain is described by the
Hamiltonian
\begin{subequations}
\label{h0}
\begin{equation}
\label{his}
\widehat{H} = - J\sum_{n=1}^{L-1}\hat\sigma_n^x\hat\sigma_{n+1}^x 
- \sum_{n=1}^{L} h_n \hat\sigma_n^z,
\end{equation}
where ${\hat\sigma_n^i}$ are the Pauli matrices corresponding to a
spin 1/2 residing on the site $i$. Using the Jordan-Wigner
transformation \cite{JW,Shura}, this model can be mapped onto a
variant of the Kitaev chain \cite{Shura,Kitaev,Kitaev08,Tsvelik}
\begin{eqnarray}
\label{hkit}
&&
\widehat{H} = -J\!\sum_{n=1}^{L-1} \!
\left(\hat c^\dagger_n \!-\! \hat c_n\right)\!
\left(\hat c^\dagger_{n+1} \!+\! \hat c_{n+1} \right)
\\
&&
\nonumber\\
&&
\qquad\qquad\qquad
-2\sum_{n=1}^{L} h_n \left(\hat c^\dagger_n \!-\! \hat c_n\right)\!
\left(\hat c^\dagger_{n} \!+\! \hat c_{n} \right).
\nonumber
\end{eqnarray}
\end{subequations}
The model originally considered by Kitaev \cite{Kitaev} maps onto the
variant of the quantum Ising model containing also the
$\hat\sigma_n^y\hat\sigma_{n+1}^y$ couplings (the XY model in a
transverse field \cite{Lieb,Suzuki}). However, it is well known
\cite{Tsvelik,Suzuki} that as long as the exchange constants in the
$xx$ and $yy$ terms are not identical, the two models are in the same
universality class. The model (\ref{hkit}) exhibits all of the essential
features of the original Kitaev chain and constitute representative
models for studies of the Majorana zero modes \cite{Kitaev08}.

\begin{figure}[t]
\centerline{\includegraphics[width=0.95\columnwidth]{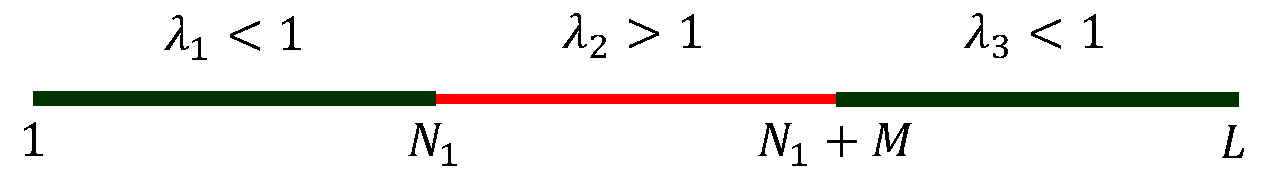}}
\caption{The nonuniform Ising-Kitaev chain split into two
  ``topological'' (dark green) and one ``trivial'' (red) regions. The
  first ``topological'' region is characterized by the parameter
  ${\lambda_1\!<\!1}$ and occupies the left part of the chain,
  ${1\!\leqslant\!{n}\!\leqslant\!{N}_1}$. The next $M$ sites are occupied by
  the ``trivial'' phase with ${\lambda_2\!>\!1}$. The remaining portion of
  the chain of the length ${N_2=L\!-\!N_1\!-\!M}$ is occupied by the second
  ``topological'' region with ${\lambda_3\!\!<1}$.}
\label{fig:model}
\end{figure}

In this paper I focus on the minimal model supporting effective
couplings between Majorana zero modes choosing the applied field $h_n$
to be piece-wise uniform (see Fig.~\ref{fig:model} for illustration)
\begin{equation}
\label{hn}
h_n=
\begin{cases}
h_1\!<\!J, & 1\leqslant n \leqslant N_1, \cr
h_2\!>\!J, & N_1\!+\!1 \leqslant n \leqslant N_1 \!+\! M, \cr
h_3\!<\!J, & N_1 \!+\! M \!+\! 1 \leqslant n \leqslant L.
\end{cases}
\end{equation}
In this case, the chain is split into three regions such that the two
``topological'' regions (of the length $N_1$ and ${N_2\!=\!L\!-\!N_1\!-\!M}$) are
separated by the ``trivial'' region of the length $M$. Since physical
properties of the model are determined by the ratios of the applied
fields to the exchange coupling $J$, it is convenient to factor out
the exchange constant $J$ introducing the parameters
\begin{equation}
\label{l}
\lambda_i = h_i/J, \qquad \lambda_1, \lambda_3 < 1, \quad \lambda_2>1.
\end{equation}

The finite-size, open-ended lattice model (\ref{h0}) is exactly
solvable. The diagonal form of the Hamiltonian (\ref{hkit}) is given
by
\begin{equation}
\label{hres1}
\widehat H = 2J\!\sum_{k=1}^L {\cal E}_k \hat \eta_k^\dagger\hat \eta_k
- J \!\sum_{k=1}^L {\cal E}_k-J\!\sum_{n=1}^L\lambda_n,
\end{equation}
where the first term describes the excitation spectrum of the model in
terms of free fermion operators $\hat\eta_k$ and the two remaining
terms yield the ground state energy. For an arbitrary choice of
$\lambda_n$, the energies ${\cal E}_k$ can be found numerically with
arbitrary precision. In the specific case (\ref{hn}), the model can
also be solved analytically. Below I present the results of the
analytic solution and compare them to the numerical results.

\subsection{Diagonalization of the Ising-Kitaev Hamiltonian}

Any quadratic Hamiltonian, including Eq.~(\ref{hkit}), can be written
in the following form (where $A_{ij}$ and $B_{ij}$ are symmetric and
antisymmetric $L\times L$ matrices, respectively)
\begin{equation}
\label{h3}
\widehat H = -2J\sum_{i,j}\left[
\hat c^\dagger_i A_{ij} \hat c_i 
+ \frac{1}{2} \left(\hat c^\dagger_i B_{ij}\hat c^\dagger_j + h.c.\right)\right].
\end{equation}
The Hamiltonian (\ref{h3}) can be diagonalized exactly using the
method \cite{Suzuki} suggested by Lieb, Schultz, and Mattis for the 1D
XY model \cite{Lieb} and used by Pfeuty to solve the uniform quantum
Ising model \cite{Pfeuty} (${h_n\!=\!h}$). The method is well known in
the theory of superconductivity \cite{Bog} and is based on the
Bogolyubov transformation
\begin{equation}
\label{btr}
\hat \eta_k \!=\! \sum_n\!\left(g_{kn} \hat c_n \!+\! h_{kn} \hat c_n^\dagger\right)\!,
\quad
\hat \eta_k^\dagger \!=\! \sum_n\!\left(g_{kn} \hat c_n^\dagger \!+\! h_{kn} \hat c_n\right)\!,
\end{equation}
where $g_{kn}$ and $h_{kn}$ are real coefficients and the resulting
operators $\hat\eta_k$ satisfy fermionic commutation relations. The
latter requirement leads to the fact that the coefficients $g_{kn}$
and $h_{kn}$ form a complete, orthonormal basis in the $L$-dimensional
Euclidean vector space.

The coefficients $g_{kn}$ and $h_{kn}$ of the Bogolyubov
transformation (\ref{btr}) can be found by assuming the diagonal form
of the Hamiltonian in terms of the operators $\hat\eta_k$ and using
the commutation relations
\[
\left[ \hat\eta_k, \widehat{H} \right] = {\cal E}_k \hat\eta_k.
\]
Using the explicit expressions (\ref{h3}) and (\ref{btr}), one finds
for the following relations for the linear combinations of $g_{kn}$
and $h_{kn}$
\begin{subequations}
  \label{ab}
  \begin{eqnarray}
&&
    {\cal E}_k\alpha_{kj} = -\sum_i\beta_{ki}(A_{ij}+B_{ij}), \nonumber\\
&&
\nonumber\\
&&
    {\cal E}_k\beta_{kj} = -\sum_i\alpha_{ki}(A_{ij}-B_{ij}),  \\
&&
\nonumber\\
&&
    {\cal E}_k^2\alpha_{kj} = \sum_i \alpha_{ki} (A-B)(A+B)_{ij},
\nonumber
  \end{eqnarray}
  where
\begin{equation}
\alpha_{kn} \!=\! g_{kn}\!+\!h_{kn}, \qquad
\beta_{kn} \!=\! g_{kn}\!-\!h_{kn}.
\end{equation}
\end{subequations}
The coefficients $\alpha_{kj}$ form the eigenvectors of the real,
symmetric matrix ${(A-B)(A+B)}$ and hence can be chosen to form a
real, orthonormal set. Then for nonzero eigenvalues, ${{\cal
    E}_k\ne0}$, the coefficients $\beta_{kj}$ are normalized
automatically. For ${{\cal E}_k=0}$, one can normalize $\beta_{kj}$.
The resulting creation operators of the Bogolyubov fermions are given
by
\begin{equation}
\label{eta1ab}
\hat\eta^\dagger_k \!=\! 
\frac{1}{2}\sum_{n=1}^{L}\left[\alpha_{kn}\left(\hat{c}^\dagger_n\!+\!\hat{c}_n\right) 
\!+\! \beta_{kn}\left(\hat{c}^\dagger_n\!-\!\hat{c}_n\right)\right].
\end{equation}
In the particular case of the Hamiltonian (\ref{hkit}), the resulting
diagonal form is given by Eq.~(\ref{hres1}).

\begin{figure}[t]
\centerline{\includegraphics[width=0.95\columnwidth]{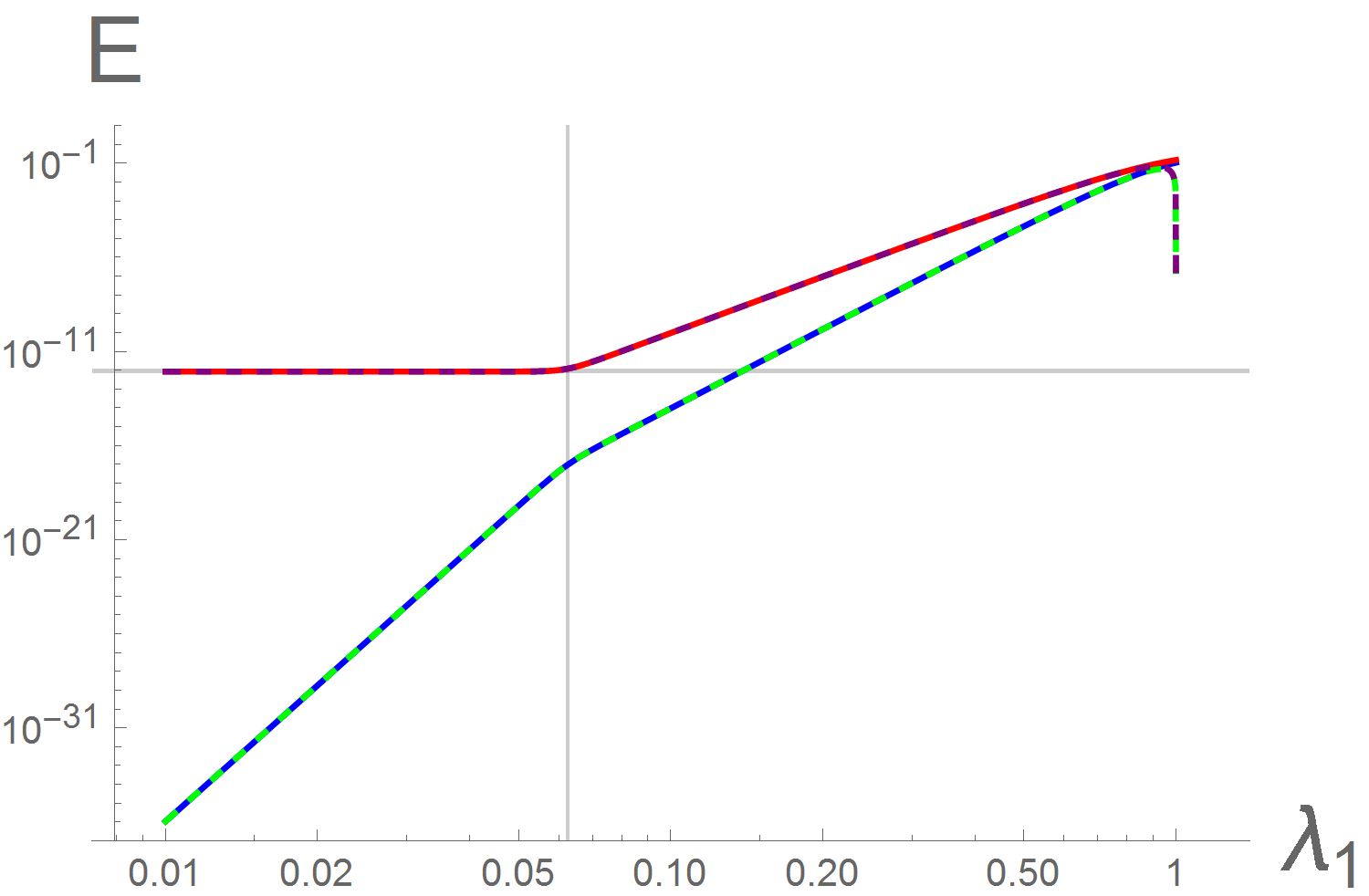}}
\caption{Energy eigenvalues ${\cal E}_{1,2}$ of the two lowest-lying
  excited states of the Ising-Kitaev chain (\ref{h0}) in
  the piece-wise uniform applied field (\ref{hn}) as a function of
  ${\lambda_1\!=\!\lambda_3}$ with ${\lambda_2\!=\!4}$ for
  ${N_1\!=\!10}$, ${M\!=\!20}$, and ${N_2\!=\!14}$. The solid curves
  represent the result of the exact numerical diagonalization. The
  dashed lines represent the analytic solutions to Eqs.~(\ref{e2m})
  and (\ref{eqth2m}). The vertical grid line corresponds to
  ${\lambda_1^{2N_1}\!=\!\lambda_2^{\!-\!2M}}$, in this particular
  case, ${\lambda_1\!=\!1/16}$. On the right side of this line the
  green dashed line corresponds to Eq.~(\ref{eres2}) and the purple --
  to Eq.~(\ref{eres1}). On the left side the green dashed line
  represents Eq.~(\ref{eres4}), the purple -- Eq.~(\ref{eres3}). The
  horizontal grid line corresponds to
  ${E=\lambda_2^{\!-\!M}\!=\!4^{-20}}$.}
\label{fig:E_log_10_14}
\end{figure}

The outlined diagonalization procedure, as well as Eq.~(\ref{hres1}),
is applicable to any quadratic Hamiltonian in any dimensionality and
allows for an efficient numerical calculation of the eigenvalues
${\cal E}_k$ and the coefficients $\alpha_{kj}$ and $\beta_{kj}$ with
arbitrary precision. However, analytic solution is manageable only in
a few relatively simple cases. Fortunately, the model (\ref{hkit})
with the specific choice (\ref{hn}) of the applied fields is one of
them. The exact single-particle energies ${\cal E}_k$ of this model
can be expressed as
\begin{eqnarray}
\label{e2m}
&&
{\cal E}^2 = 1 + \lambda_1^2 + 2\lambda_1\cos\vartheta_1 
\\
&&
\nonumber\\
&&
\qquad
=
1 + \lambda_2^2 + 2\lambda_2\cos\vartheta_2
=
1 + \lambda_3^2 + 2\lambda_3\cos\vartheta_3,
\nonumber
\end{eqnarray}
in terms of nontrivial solutions, $\vartheta_i$, to the equation
\begin{subequations}
\label{eqth2m}
\begin{eqnarray}
\lambda_1 D_1(N_1) D_2(M, N_2) \!=\! \lambda_2 D_1(N_1\!+\!1) D_2(M\!+\!1, N_2),
\quad
\end{eqnarray}
where
\begin{eqnarray}
&&
D_1(N_1)\!=\! \lambda_1\sin N_1\vartheta_1 \!+\! \sin(N_1\!-\!1)\vartheta_1,
\\
&&
\nonumber\\
&&
D_2(M, N_2) \!=\! \lambda_3 \sin(N_2\!+\!1)\vartheta_3 \sin M \vartheta_2 
\\
&&
\nonumber\\
&&
\qquad\qquad\qquad\qquad
-\lambda_2 \sin N_2\vartheta_3 \sin (M\!-\!1) \vartheta_2.
\nonumber
\end{eqnarray}
\end{subequations}
The latter equalities in Eq.~(\ref{e2m}) provide additional
constraints on $\vartheta_i$, which guarantee the uniqueness of the
solution.

Consider now the energy (\ref{e2m}) as a function of $\vartheta_i$,
regardless of which values of $\vartheta_i$ are allowed by
Eq.~(\ref{eqth2m}). For real $\vartheta_i$, this function exhibits a
minimum at ${\vartheta_i=\pi}$. The minimum value of the energy
gives a reasonable lower bound for the bulk gap
\begin{equation}
\label{gap}
\Delta\approx 2J\min_{i=1,2,3} |1-\lambda_i|.
\end{equation}
Hence, any subgap states including nearly zero-energy states are
described by complex solutions to Eq.~(\ref{eqth2m}).

\subsection{Nearly degenerate ground states}

The Ising-Kitaev chain split into two ``topological'' and one
``trivial'' region possesses two single-particle excitations (hereafter
denoted by ${k\!=\!1,2}$) that are nearly degenerate with the ground
state. As long as the parameters $\lambda_i$ are not too close to
unity and the sizes of the three regions are not too small, such that
the three quantities $\lambda_1^{2N_1}$, $\lambda_3^{2N_2}$, and
$\lambda_2^{-2M}$ are exponentially small, the energies and the
wavefunctions of these states can be found analytically. Already the
leading-order expression shows excellent agreement with the exact
numerical diagonalization of the model as illustrated in
Figs.~\ref{fig:E_log_10_14} and \ref{fig:E_log_10_10}. The visible
discrepancy between the analytic and numerical results for
${\lambda_1\sim1}$ is to be expected: there the above parameters cease
being exponentially small and the approximate analytic expressions
become invalid.

Without specifying the relation between the three exponentially small
parameters, even the leading-order expression for the two eigenvalues
${\cal E}_{1,2}$ is rather cumbersome. Therefore here I focus on two
representative limiting cases (the symbolic expression
${{\cal O}(\lambda^{N})}$ hereafter denotes the omitted subleading,
exponentially small terms).

\subsubsection{Asymmetric chain}

If the two ``topological'' regions of the chain are not symmetric,
then compact expressions for the energies ${\cal E}_{1,2}$ can be
found under following assumptions.
\begin{subequations}
\label{lex}

(i) {\it ``Strong barrier''}. If
${\lambda_1^{2N_1}>\lambda_3^{2N_2}\gg\lambda_2^{\!-\!2M}}$, the two
nearly zero-energy states are determined by the two ``topological''
regions of the chain, independently of the size of the ``trivial''
region. The first excited state has the energy
\begin{equation}
\label{eres2}
{\cal E}_1 = (1\!-\!\lambda^2_3) 
\sqrt{\frac{\lambda_2^2\!-\!1}{\lambda_2^2\!-\!\lambda^2_3}}\lambda_3^{N_2}
+{\cal O}(\lambda^{N}),
\end{equation}
while the energy of the second excited state is 
\begin{equation}
\label{eres1}
{\cal E}_2 = (1\!-\!\lambda^2_1) 
\sqrt{\frac{\lambda_2^2\!-\!1}{\lambda_2^2\!-\!\lambda^2_1}}\lambda_1^{N_1}
+{\cal O}(\lambda^{N}),
\end{equation}
These results are illustrated in Fig.~\ref{fig:E_log_10_14} by the
dashed lines to the right of the vertical grid line (marking the end
of the above parameter region
${\lambda_1^{2N_1}=\lambda_2^{\!-\!2M}}$). Vanishing of the energies
(\ref{eres2}) and (\ref{eres1}) at the point ${\lambda_1=\lambda_3=1}$
is the artifact of the approximation. As the parameters $\lambda_i$
approach unity, the approximate expressions reported here become
invalid (while it is possible to write down exact expressions for
${\cal E}_{1,2}$ that are valid also near the critical point, their
algebraic complexity renders them practically useless).

(ii) {\it ``Weak barrier''}.  In the case
${\lambda_1^{2N_1},\lambda_3^{2N_2}\ll\lambda_2^{\!-\!2M}}$, the
larger eigenvalue ${\cal E}_2$ is determined by the ``trivial''
region of the chain
\begin{equation}
\label{eres3}
{\cal E}_2 = (\lambda_2^2\!-\!1) 
\sqrt{\frac{(1\!-\!\lambda^2_1)(1\!-\!\lambda^2_3)}
{(\lambda_2^2\!-\!\lambda^2_1)(\lambda_2^2\!-\!\lambda^2_3)}}\;\lambda_2^{-M}
+{\cal O}(\lambda^{N}),
\end{equation}
while the energy of the lowest excited state is determined by the
two ``topological'' regions combined 
\begin{equation}
\label{eres4}
{\cal E}_1 =
\sqrt{(1\!-\!\lambda^2_1)(1\!-\!\lambda^2_3)}\;
\lambda_1^{N_1}\lambda_3^{N_2}\lambda_2^{M}+{\cal O}(\lambda^{N}).
\end{equation}
These results are illustrated in Fig.~\ref{fig:E_log_10_14} by the
dashed lines on the left side of the vertical grid line.

\end{subequations}

\subsubsection{Symmetric chain}

In the symmetric case, ${\lambda_1^{N_1}=\lambda_3^{N_2}}$, the two
energies (\ref{eres1}) and (\ref{eres2}) coincide. In this case, one
has to consider the subleading terms neglected so far, as the
eigenvalues of the finite-size chain (\ref{hkit}) are never truly
degenerate.

\begin{figure}[t]
\centerline{\includegraphics[width=0.95\columnwidth]{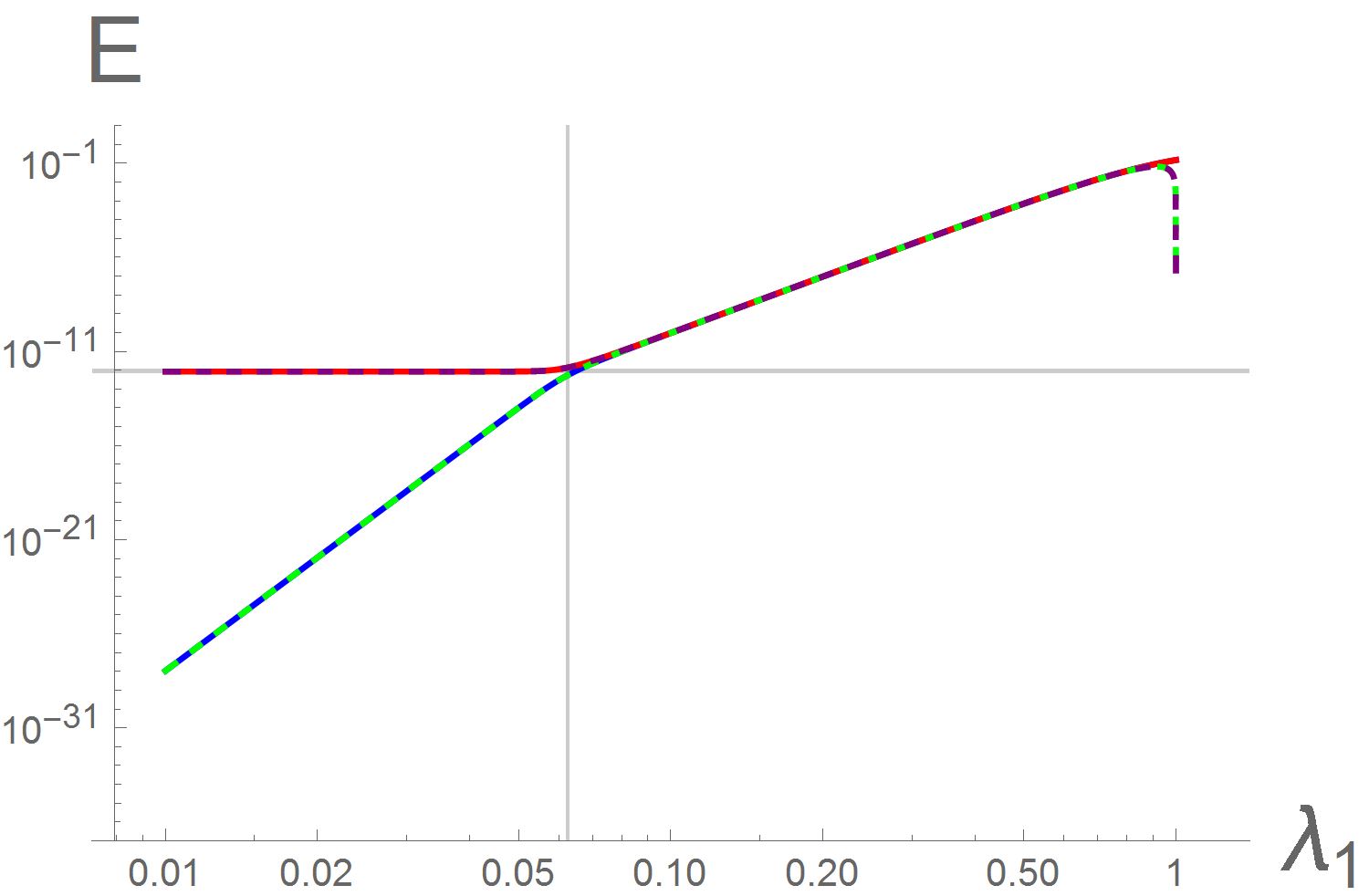}}
\caption{Energy eigenvalues ${\cal E}_{1,2}$ of the two lowest-lying
  excited states of the symmetric Ising-Kitaev chain (\ref{h0}) in the
  piece-wise uniform applied field (\ref{hn}) as a function of
  ${\lambda_1\!=\!\lambda_3}$ with ${\lambda_2\!=\!4}$ for
  ${N_1\!=\!N_2\!=\!10}$ and ${M\!=\!20}$. The solid curves represent
  the result of the exact numerical diagonalization. The dashed lines
  represent the analytic solutions to Eqs.~(\ref{e2m}) and
  (\ref{eqth2m}). The vertical grid line corresponds to
  ${\lambda_1^{2N_1}\!=\!\lambda_2^{\!-\!2M}}$, in this particular
  case, ${\lambda_1\!=\!1/16}$. On the right side of this line the
  green and purple dashed lines corresponds to the two eigenvalues in
  Eq.~(\ref{xsym}). On the left side the green dashed line represents
  Eq.~(\ref{eres4}), the purple -- Eq.~(\ref{eres3}). The horizontal
  grid line corresponds to ${E\!=\!\lambda_2^{\!-\!M}\!=\!4^{-20}}$.}
\label{fig:E_log_10_10}
\end{figure}

(i) {\it ``Strong barrier''}. Assuming
${\lambda_1^{2N_1}\gg\lambda_2^{\!-\!2M}}$, the resulting energies are
given by
\begin{eqnarray}
\label{xsym}
&&
{\cal E}_{1(2)}^{\rm sym} =
(1\!-\!\lambda^2_1) 
\sqrt{\frac{\lambda_2^2\!-\!1}{\lambda_2^2\!-\!\lambda^2_1}}\lambda_1^{N_1}
\\
&&
\nonumber\\
&&
\qquad\qquad
\times
\left[
1\mp\frac{1}{2}
\lambda_1^{-N_1}\lambda_2^{-M}\!
\sqrt{\frac{\lambda_2^2\!-\!1}{\lambda_2^2\!-\!\lambda^2_1}}\right]
+{\cal O}(\lambda^{N}).
\nonumber
\end{eqnarray}
This result is illustrated in Fig.~\ref{fig:E_log_10_10} to the right
of the vertical grid line (the exponentially small difference between
the two energies (\ref{xsym}) is indistinguishable on the scale of the
plot).

(ii) {\it ``Weak barrier''}. In the limit,
${\lambda_1^{2N_1}\ll\lambda_2^{\!-\!2M}}$, no spurious degeneracy
occurs and hence the expressions (\ref{eres3}) and (\ref{eres4}) are
still valid, see Fig.~\ref{fig:E_log_10_10} (to the left of the
vertical grid line).

\section{Majorana zero modes}

Elementary excitations of the model can be interpreted in terms of
Majorana fermions \cite{Shura,Kitaev}. In fact, the fermionic form
(\ref{hkit}) of the Hamiltonian is already written in terms of the
lattice Majorana fermions \cite{Shura}
\begin{equation}
\label{maj}
\hat\zeta_n \!=\! \hat c^\dagger_n \!+\! \hat c_n, \qquad
\hat\xi_n \!=\! -i\left(\hat c^\dagger_n \!-\! \hat c_n\right).
\end{equation}
In terms of the operators (\ref{maj}), the creation operator,
$\hat\eta^\dagger_k$, of a single-particle excitation has the form
[cf. Eq.~(\ref{eta1ab})]
\begin{equation}
\label{eta1mf}
\hat\eta^\dagger_k \!=\! \frac{1}{2}\sum_{n=1}^{L}\left[\alpha_{kn}\hat{\zeta}_n 
\!+\! i\beta_{kn}\hat{\xi}_n\right].
\end{equation}
The two linear combinations
\begin{equation}
\label{gamma}
\hat\gamma^{(\alpha)}_k\!=\! \sum_{n=1}^{L}\alpha_{kn}\hat{\zeta}_n, \qquad
\hat\gamma^{(\beta)}_k \!=\! \sum_{n=1}^{L}\beta_{kn}\hat{\xi}_n,
\end{equation}
are themselves Majorana operators \cite{Kitaev} in the sense that they
satisfy the Majorana commutation relations
\begin{equation}
\label{gcom}
\left\{\hat\gamma^{(\alpha)}_k, \hat\gamma^{(\beta)}_k\right\}=0, \qquad
\left(\hat\gamma^{(\alpha)}_k\right)^2 \!=\left(\hat\gamma^{(\beta)}_k\right)^2 \!=1.
\end{equation}
The latter property follows from the fact that the vectors
$\alpha_{kn}$ and $\beta_{kn}$ are normalized and mutually orthogonal.

By definition, the Majorana operators (\ref{gamma}) are nonlocal
linear combinations \cite{Kitaev} of the more conventional
\cite{Shura} Majorana fermions (\ref{maj}). Typically, these
combinations involve {\it all} sites of the chain
\cite{Kitaev}. However, for the two lowest excited states
(\ref{eres1})-(\ref{xsym}) the amplitudes $\alpha_{1(2)n}$ and
$\beta_{1(2)n}$ exhibit the exponential decay away from the interface
points of the chain, allowing one to treat the nearly zero-energy
Majorana states $\hat\gamma^{(\alpha)}_{1(2)}$ and
$\hat\gamma^{(\beta)}_{1(2)}$ as essentially localized \cite{Kitaev},
see Figs.~\ref{fig:ab_10_14} and \ref{fig:ab_10_10}.

\subsection{Asymmetric chain}

\begin{figure}[t]
\centerline{\includegraphics[width=0.99\columnwidth]{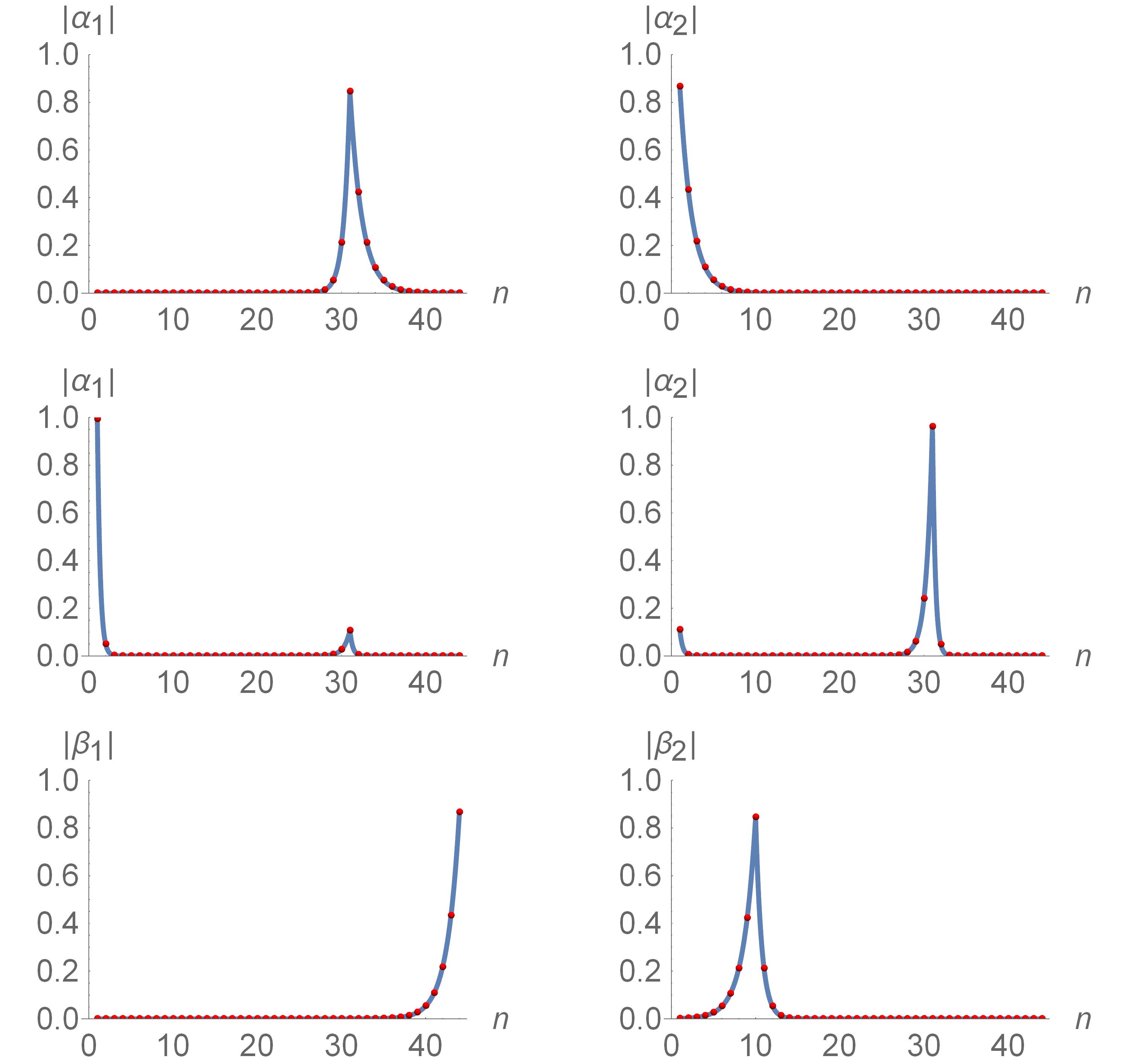}}
\caption{Majorana amplitudes of the two nearly zero-energy eigenstates
  of the Ising-Kitaev chain (\ref{h0}) with ${N_1\!=\!10}$,
  ${M\!=\!20}$, and ${N_2\!=\!14}$ in the piece-wise uniform applied
  field (\ref{hn}). The red dots represent the result of the exact
  numerical diagonalization. The curves represent the analytic
  solutions. Top row: the amplitudes $|\alpha_{1n}|$ and
  $|\alpha_{1n}|$ in the strong barrier case,
  ${\lambda_1\!=\!\lambda_3\!=\!1/2}$, ${\lambda_2\!=\!4}$. The curves
  are given in Eqs.~(\ref{a1r}) and (\ref{a2r}). Middle row: the
  amplitudes $|\alpha_{1n}|$ and $|\alpha_{1n}|$ in the weak barrier
  case, ${\lambda_1\!=\!\lambda_3\!=\!1/20}$, ${\lambda_2\!=\!4}$,
  exhibiting weak delocalization. The curves are given in
  Eqs.~(\ref{a3r}) and (\ref{a4r}). Bottom row: the amplitudes
  $|\beta_{1n}|$ and $|\beta_{1n}|$ for
  ${\lambda_1\!=\!\lambda_3\!=\!1/2}$, ${\lambda_2\!=\!4}$. The curves
  are given by either Eqs.~(\ref{a1r}) and (\ref{a2r}) or
  Eqs.~(\ref{a3r}) and (\ref{a4r}).}
\label{fig:ab_10_14}
\end{figure}

(i) {\it ``Strong barrier''}. In the limit
${\lambda_1^{2N_1}>\lambda_3^{2N_2}\gg\lambda_2^{\!-\!2M}}$, the
leading behavior of the energy eigenvalues is given by
Eqs.~(\ref{eres2}) and (\ref{eres1}). The corresponding amplitudes
$\alpha_{1(2)n}$ and $\beta_{1(2)n}$ can also be written in compact
from, again retaining only the leading exponential terms. The first
excited state (\ref{eres2}) is characterized by the amplitudes
\begin{eqnarray}
&&
\alpha_{1n} \!=\! (\!-\!1)^{n}
\begin{cases}
{\cal O}(\lambda^{N}), & \!\!1\!\leqslant\! n \!\leqslant\! N_1, \cr
c_3
\lambda_2^{n\!-\!1\!-\!N_1\!-\!M}, 
& \!\!1\!\leqslant\! n\!-\!N_1 \!\leqslant\! M, \cr
c_3
\lambda_3^{n\!-\!1\!-\!N_1\!-\!M},
& \!\!1\!+\!N_1\!+\!M\!\leqslant\! n \!\leqslant\! L,
\end{cases}
\nonumber\\
&&
\label{a1r}
\\
&&
\beta_{1n} \!=\! (\!-\!1)^{n\!+\!1}\! 
\begin{cases}
{\cal O}(\lambda^{N}),  & \!\!\!1\!\leqslant\! n \!\leqslant\! N_1\!+\!M, \cr
s_3\lambda_3^{L\!-\!n}, 
& \!\!\!1\!+\!N_1\!+\!M\!\leqslant\! n \!\leqslant\! L.
\end{cases}
\nonumber
\end{eqnarray}
where the symbolic expression ${{\cal O}(\lambda^{N})}$ denoting the
subleading terms is omitted in some lines for brevity and
\begin{equation}
c_j=\sqrt{\frac{(1\!-\!\lambda_j^2)(\lambda_2^2\!-\!1)}{\lambda_2^2\!-\!\lambda_j^2}},
\qquad
s_j=\sqrt{1\!-\!\lambda_j^2}.
\end{equation}
Hence with exponential accuracy, the lowest-energy excitation of the
model can be described by the single fermion, $\hat\eta_1$, confined
to the second ``topological''region of the chain,
cf. Eq.~(\ref{eta1mf}). The amplitudes (\ref{a1r}) are illustrated in
the two top panels in Fig.~\ref{fig:ab_10_14}. The spread of the
localized Majorana states over several lattice sites exhibited by
Eq.~(\ref{a1r}) is a generic feature \cite{Kitaev} that can be seen
also in the continuum limit \cite{Halperin}.

The second eigenvalue (\ref{eres1}) is characterized by the amplitudes
\begin{eqnarray}
&&
\alpha_{2n} \!=\! (\!-\!1)^{n\!-\!1}\!
\begin{cases}
s_1\lambda_1^{n\!-\!1}\!+\!{\cal O}(\lambda^{N}), 
& \!\!\!1\!\leqslant\! n \!\leqslant\! N_1, \cr
{\cal O}(\lambda^{N}), & \!\!\!1\!\leqslant\! n\!-\!N_1 \!\leqslant\! L,
\end{cases}
\nonumber\\
&&
\label{a2r}
\\
&&
\beta_{2n} \!=\! (\!-\!1)^{n}
\begin{cases}
c_1\lambda_1^{N_1\!-\!n}\!+\!{\cal O}(\lambda^{N}),
& \!\!1\!\leqslant\! n \!\leqslant\! N_1, \cr
c_1
\lambda_2^{N_1\!-\!n}\!+\!{\cal O}(\lambda^{N}), 
& \!\!1\!\leqslant\! n\!-\!N_1 \!\leqslant\! M, \cr
{\cal O}(\lambda^{N}),
& \!\!1\!+\!N_1\!+\!M\!\leqslant\! n \!\leqslant\! L.
\end{cases}
\nonumber
\end{eqnarray}
The corresponding excitation $\hat\eta_2$ is confined to the
first ``topological''region of the chain. The amplitudes (\ref{a2r})
are illustrated in the two bottom panels in Fig.~\ref{fig:ab_10_14}.

The results (\ref{a1r}) and (\ref{a2r}) confirm that in the limit
${\lambda_1^{2N_1}>\lambda_3^{2N_2}\gg\lambda_2^{\!-\!2M}}$ the two
lowest-energy excitations of the model behave similarly to those of
the two independent ``topological'' regions. In particular, they
exhibit four nearly zero-energy Majorana fermions localized at the
edges of the ``topological'' regions.

(ii) {\it ``Weak barrier''}. In the limit,
${\lambda_1^{2N_1},\lambda_3^{2N_2}\!\ll\!\lambda_2^{\!-\!2M}}$, the
structure of the wave-functions of the two lowest excited states is
significantly different. The first excited state (\ref{eres2}) is
characterized by the amplitudes
\begin{eqnarray}
&&
\!\!\!\!\alpha_{1n} \!=\! (\!-\!1)^{n} s_1
\begin{cases}
\lambda_1^{n-1}, & \!\!1\!\leqslant\! n \!\leqslant\! N_1, \cr
\lambda_1^{N_1}
\lambda_2^{n\!-\!1\!-\!N_1}, 
& \!\!1\!\leqslant\! n\!-\!N_1 \!\leqslant\! M, \cr
\lambda_1^{N_1}\lambda_2^{M}
\lambda_3^{n\!-\!1\!-\!N_1\!-\!M},
& \!\!1\!+\!N_1\!+\!M\!\leqslant\! n \!\leqslant\! L,
\end{cases}
\nonumber\\
&&
\label{a3r}
\\
&&
\!\!\!\!\beta_{1n} \!=\! (\!-\!1)^{n\!+\!1}\! 
\begin{cases}
{\cal O}(\lambda^{N}),  & \!\!\!1\!\leqslant\! n \!\leqslant\! N_1\!+\!M, \cr
s_3\lambda_3^{L\!-\!n}, 
& \!\!\!1\!+\!N_1\!+\!M\!\leqslant\! n \!\leqslant\! L,
\end{cases}
\nonumber
\end{eqnarray}
The amplitude $\beta_{1n}$ is identical with Eq.~(\ref{a1r}), but the
amplitude $\alpha_{1n}$ has changed. In the first ``topological''
region of the chain, it behaves as the corresponding amplitude of the
second excited state (\ref{a2r}) of the strong barrier case. Moreover,
there is a nonzero probability to find this quasiparticle also at the
interface between the ``trivial'' and the second ``topological''
regions, see Fig.~\ref{fig:ab_10_14}, i.e. the corresponding Majorana
fermion is essentially ``delocalized'' between the two points.

The ``delocalization'' of the Majorana amplitude $\alpha_{1n}$ in
Eq.~(\ref{a3r}) is rather weak. For the following choice of values of
the parameters (\ref{l}), $\lambda_1\!=\!\lambda_3\!=\!1/20$,
$\lambda_2\!=\!4$, and the sizes of the chain segments $N_1\!=\!10$,
$N_2\!=\!14$, $M\!=\!20$, the peak values of the amplitude
$\alpha_{1n}$ are $|\alpha_{1,1}|\!=\!0.999$ and
$|\alpha_{1,31}|\!=\!0.107$. Whether this feature survives the
thermodynamic limit depends on what happen to the value of the product
of a small and large parameters ${\lambda_1^{N_1}\lambda_2^{M}}$ in
the limiting procedure.

The second eigenvalue (\ref{eres1}) is characterized by the amplitudes
\begin{eqnarray}
&&
\!\!\!\!\alpha_{2n} \!=\! (\!-\!1)^{n}\!
\begin{cases}
-(s_1^2/c_3)\lambda_2^M\lambda_1^{N_1\!+\!n\!-\!1}, & \!\!1\!\leqslant\! n \!\leqslant\! N_1, \cr
c_3
\lambda_2^{n\!-\!1\!-\!N_1\!-\!M}, 
& \!\!1\!\leqslant\! n\!-\!N_1 \!\leqslant\! M, \cr
c_3
\lambda_3^{n\!-\!1\!-\!N_1\!-\!M},
& \!\!1\!+\!N_1\!+\!M\!\leqslant\! n \!\leqslant\! L,
\end{cases}
\nonumber\\
&&
\label{a4r}
\\
&&
\!\!\!\!\beta_{2n} \!=\! (\!-\!1)^{n}
\begin{cases}
c_1\lambda_1^{N_1\!-\!n},
& \!\!1\!\leqslant\! n \!\leqslant\! N_1, \cr
c_1
\lambda_2^{N_1\!-\!n}, 
& \!\!1\!\leqslant\! n\!-\!N_1 \!\leqslant\! M, \cr
{\cal O}(\lambda^{N}),
& \!\!1\!+\!N_1\!+\!M\!\leqslant\! n \!\leqslant\! L.
\end{cases}
\nonumber
\end{eqnarray}
Similarly to the previous case, the amplitude $\beta_{2n}$ remains the
same as in Eq.~(\ref{a2r}), while the amplitude $\alpha_{1n}$ exhibits
the weak ``delocalization''. For the same choice of parameters
($\lambda_1\!=\!\lambda_3\!=\!1/20$, $\lambda_2\!=\!4$ and
$N_1\!=\!10$, $N_2\!=\!14$, $M\!=\!20$), I find
$|\alpha_{2,1}|\!=\!0.967$ and $|\alpha_{2,31}|\!=\!0.111$.

In contrast to the strong barrier case, the wave-function of
lowest-energy fermion $\hat\eta_1$ is now mostly spread between the
two outer edges of the chain, with a small weight at the interface
between the ``trivial'' and the second ``topological'' region. The
second excitation $\hat\eta_2$ is mostly confined to the edges of the
``trivial'' region, with the small weight at the beginning of the
chain, see Fig.~\ref{fig:ab_10_14}.

\subsection{Symmetric chain}

\begin{figure}[t]
\centerline{\includegraphics[width=0.99\columnwidth]{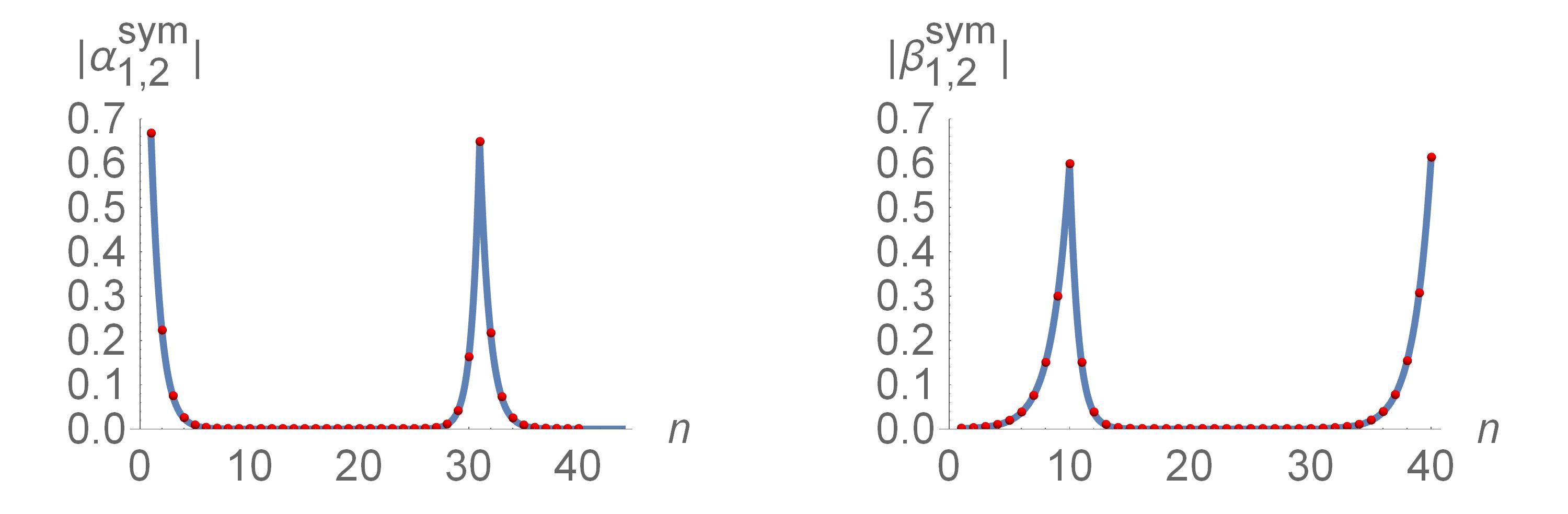}}
\caption{Majorana amplitudes (\ref{asym}) of the two nearly
  zero-energy eigenstates of the symmetric Ising-Kitaev chain
  (\ref{h0}) in the piece-wise uniform applied field (\ref{hn}) with
  ${\lambda_1\!=\!\lambda_3\!=\!1/2}$, ${\lambda_2\!=\!4}$,
  ${N_1\!=\!N_2\!=\!10}$, and ${M\!=\!20}$. The red dots represent the
  result of the exact numerical diagonalization. The curves represent
  the analytic solutions given in Eqs.~(\ref{asym}).}
\label{fig:ab_10_10}
\end{figure}

The two lowest-energy excitations of the symmetric chain with {\it the
strong barrier},
${\lambda_1^{2N_1}\!=\!\lambda_3^{2N_2}\!\gg\!\lambda_2^{\!-\!2M}}$,
are characterized by the amplitudes, see Fig.~\ref{fig:ab_10_10} (the
symbolic expression ${{\cal O}(\lambda^{N})}$ denoting the subleading
terms is omitted for brevity)
\begin{eqnarray}
&&
\alpha_{1(2),n}^{\rm sym} \!=\! \frac{(\!-\!1)^{n}}{\sqrt{2}}
\begin{cases}
\mp\!\sqrt{1\!-\!\lambda_1^2}\;\lambda_1^{n\!-\!1}, & \!\!1\!\leqslant\! n \!\leqslant\! N_1, \cr
c_1
\lambda_2^{n\!-\!1\!-\!N_1\!-\!M}, 
& \!\!1\!\leqslant\! n\!-\!N_1 \!\leqslant\! M, \cr
c_1
\lambda_1^{n\!-\!1\!-\!N_1\!-\!M},
& \!\!1\!+\!N_1\!+\!M\!\leqslant\! n \!\leqslant\! L,
\end{cases}
\nonumber\\
&&
\label{asym}
\\
&&
\beta_{1(2),n}^{\rm sym} \!=\! (\!-\!1)^{n}
\begin{cases}
\mp c_1\lambda_1^{N_1\!-\!n},
& \!\!1\!\leqslant\! n \!\leqslant\! N_1, \cr
\mp c_1
\lambda_2^{N_1\!-\!n}, 
& \!\!1\!\leqslant\! n\!-\!N_1 \!\leqslant\! M, \cr
\sqrt{1\!-\!\lambda_1^2}\;\lambda_1^{L\!-\!n},
& \!\!1\!+\!N_1\!+\!M\!\leqslant\! n \!\leqslant\! L.
\end{cases}
\nonumber
\end{eqnarray}
In this case, the excitations $\hat\eta_{1,2}^\dagger$ are no longer
confined to one of the two ``topological'' regions of the chain, but
are spread symmetrically over both of them.

In the {\it weak barrier} case, the amplitudes $\alpha_{n}$ and
$\beta_{n}$ are still described by Eqs.~(\ref{a3r}) and (\ref{a4r}).


The exponential decay of the amplitudes (\ref{a1r})-(\ref{asym}) can
be described in terms of a correlation length, which is specific to
each of the three regions of the chain
\begin{equation}
\label{cl}
\ell_{0i}\sim 1/|\ln\lambda_i|.
\end{equation} 
In experiments on discreet systems \cite{Mooij,Nori,Oreg13,Nori16},
the realistic values of $\lambda_i$ might not be extreme and hence the
correlation lengthes (\ref{cl}) might not be very small. In such case,
even the localized Majorana fermions are spread over several lattice
sites as illustrated in Figs.~\ref{fig:ab_10_14} and
\ref{fig:ab_10_10}.

\section{Effective low energy theory}

Applications to quantum computation \cite{Alicea} involve adiabatic
manipulations of the Majorana zero modes. This means that any external
perturbation applied to the system should be slow enough to avoid
exciting higher-energy gapped states. The remaining low-energy sector
of the theory consists of the ground state $|GS\rangle$ and the nearly
degenerate excitations that can be interpreted in terms of Majorana
zero modes.

For the specific model considered in this paper, the low-energy sector
contains four states
\begin{equation}
\label{les}
|GS\rangle, \qquad \eta^\dagger_1|GS\rangle, 
\qquad \eta^\dagger_2|GS\rangle, 
\qquad \eta^\dagger_1\eta^\dagger_2|GS\rangle, 
\end{equation}
where the last state is the two-particle excitation. These four states
can be further split into two groups of mutually orthogonal states,
belonging to the two parity sectors of the model where the total
fermion number is either even or odd.

Projecting the Hamiltonian (\ref{hkit}) onto either of the above
sectors, one finds the effective low-energy theory. In the one-fermion
(odd) sector, the effective Hamiltonian has the simplest form in the
basis of the Majorana fermions $\hat\gamma^{(\alpha)}_{1(2)}$ and
$\hat\gamma^{(\beta)}_{1(2)}$.


In the {\it asymmetric chain} with the {\it strong barrier}, i.e. in
the limit
${\lambda_1^{2N_1}\!>\!\lambda_3^{2N_2}\!\gg\!\lambda_2^{\!-\!2M}}$,
the localized Majorana fermions describe the exact eigenstates of the
model. Hence, the projected Hamiltonian in the basis the four Majorana
states (counted from left to right) has the block-diagonal structure
\begin{equation}
\label{he1}
\widehat{H}_{\rm eff} \propto
\begin{pmatrix}
0 & -i{\cal E}_2 & 0 & 0 \cr
i{\cal E}_2 & 0 & 0 & 0 \cr
0 & 0 & 0 & -i{\cal E}_1 \cr
0 & 0 & i{\cal E}_1 & 0
\end{pmatrix}.
\end{equation}
Note the absence of any coupling between the two pairs of the Majorana
zero modes, $\gamma_2$, $\gamma_3$ and $\gamma_1$, $\gamma_4$ (since
the exact orthogonal eigenstates of the model are described by
$\gamma_1$, $\gamma_2$ and $\gamma_3$, $\gamma_4$).

In contrast, in the case of the {\it symmetric chain} with the {\it
  strong barrier},
${\lambda_1^{2N_1}=\lambda_3^{2N_2}\gg\lambda_2^{\!-\!2M}}$, the
Majorana amplitudes (\ref{asym}) are not localized at single interface
points in the chain. One can still represent the effective Hamiltonian
in the basis of the four localized Majorana fermions. However, now
these objects are no longer associated with the exact eigenstates and
hence additional couplings appear. Introducing a short-hand notation
for the eigenvalues (\ref{xsym})
\begin{equation}
\label{esym}
{\cal E}_{1(2)}^{\rm sym} = \epsilon \pm \delta, \qquad \delta\ll\epsilon,
\end{equation}
I find the following Hamiltonian (using an obvious basis rotation)
\begin{equation}
\label{he2}
\widehat{H}_{\rm eff} \propto
\begin{pmatrix}
0 & -i\epsilon & 0 & -i\delta\cr
i\epsilon & 0 & -i\delta & 0 \cr
0 & i\delta & 0 & -i\epsilon \cr
i\delta & 0 & i\epsilon & 0
\end{pmatrix}.
\end{equation}
The Hamiltonian (\ref{he2}) is no longer block-diagonal: the two pairs
of the Majorana zero modes, $\gamma_2$, $\gamma_3$ and $\gamma_1$,
$\gamma_4$, are now coupled. The effective coupling of both pairs is
the same (given by $\delta$) despite the large difference in their
respective separation. This result is the consequence of the nonlocal
nature of the eigenstates of the model.

In the {\it weak barrier} case, the low-energy excitations are no
longer confined to the ``topological'' regions of the chain. In order
to take into account the ``delocalization'' of the amplitudes
$\alpha_{1(2)n}$, see Eqs.~(\ref{a3r}) and (\ref{a4r}), I denote the
ratio of the peak values of $\alpha_{1(2)n}$ as
${\kappa\!\approx\!\lambda_1^{N_1}\!\lambda_2^Ms1/c3\!\ll\!1}$ (up to
exponentially small corrections) and find the low-energy Hamiltonian
(in the same basis of $\gamma_i$ counted from left to right)
\begin{equation}
\label{he3}
\widehat{H}_{\rm eff} \propto
\begin{pmatrix}
0 & -i\kappa{\cal E}_2 & 0 & -i{\cal E}_1 \cr
i\kappa{\cal E}_2 & 0 & -i{\cal E}_2 & 0 \cr
0 & i{\cal E}_2 & 0 & -i\kappa{\cal E}_1 \cr
i{\cal E}_1 & 0 & i\kappa{\cal E}_1 & 0
\end{pmatrix}.
\end{equation}
In contrast to Eqs.~(\ref{he1}) and (\ref{he2}), the leading terms in
Eq.~(\ref{he3}) are confined to the antidiagonal. This results
reflects the structure of the single-fermion excitations (\ref{a3r})
and (\ref{a4r}): the lowest-energy excitation (\ref{a3r}) is mostly
spread between the two outer edges of the chain (and is described by
$\gamma_1$, $\gamma_4$), while the fermion (\ref{a4r}) -- to the edges
of the ``trivial'' region (corresponding to $\gamma_2$, $\gamma_3$).
Neglecting weak ``delocalization'' completely (i.e. in the limit
${\kappa\!\rightarrow\!0}$), the Hamiltonian (\ref{he3}) can be
made block-diagonal by renumbering the localized Majorana
operators.


\begin{figure}[t]
\centerline{\includegraphics[width=0.6\columnwidth]{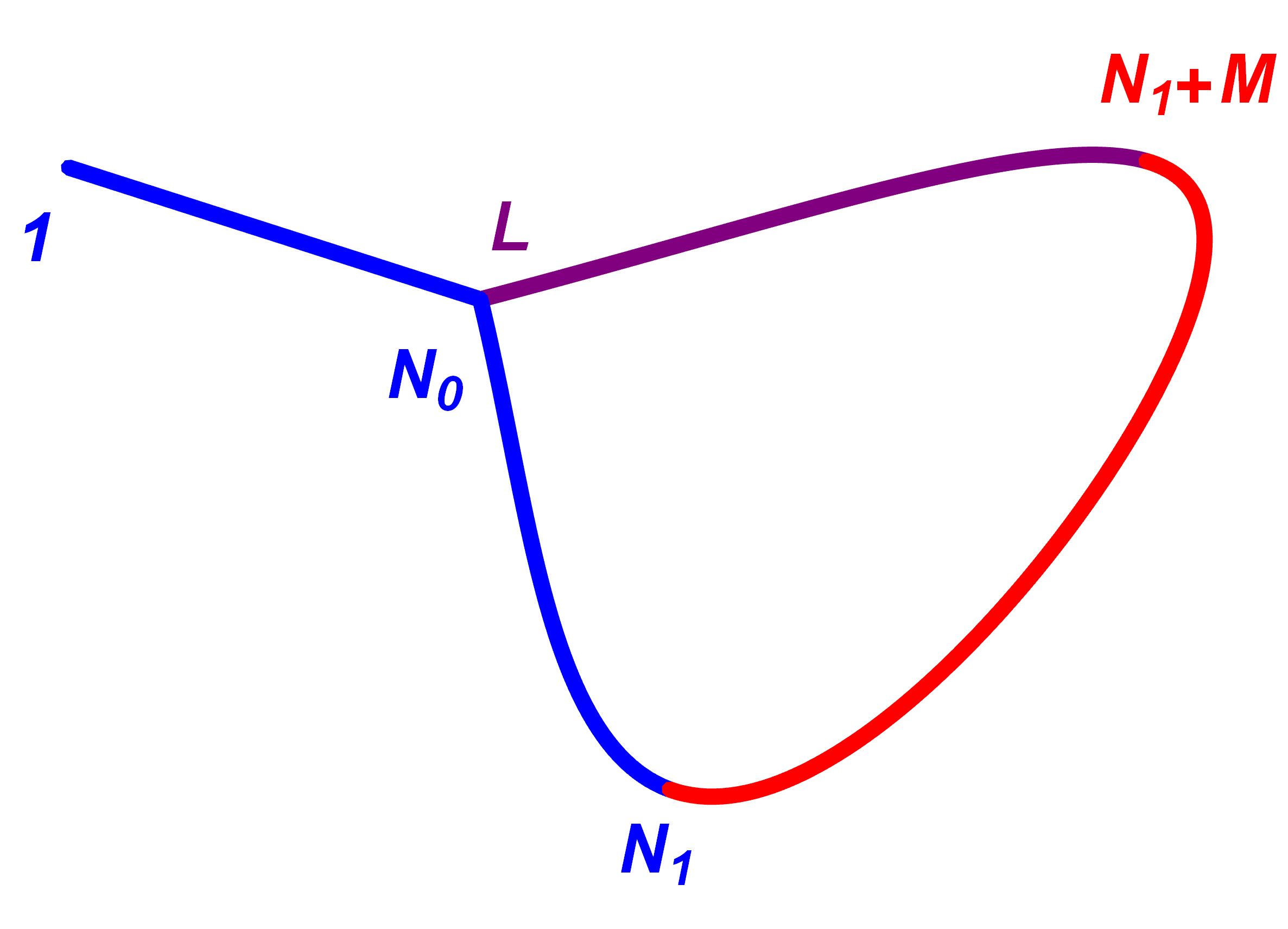}}
\caption{The nonuniform Ising-Kitaev chain with a T-junction. The
  chain contains two ``topological'' (blue and purple) regions and one
  ``trivial'' (red) region. The first ``topological'' region is
  characterized by the parameter ${\lambda_1\!<\!1}$ and contains the
  first ${{N}_1}$ sites. The next $M$ sites are occupied by the
  ``trivial'' phase with ${\lambda_2\!>\!1}$. The remaining portion of
  the chain of the length ${N_2\!=\!L\!-\!N_1\!-\!M}$ is occupied by
  the second ``topological'' region with ${\lambda_3\!<\!1}$. The
  T-junction is located at the site $N_0$. In this paper, I report
  results for the case where the T-junction is approximately in the
  middle of the first ``topological'' region of the chain.}
\label{fig:modelT}
\end{figure}

\section{Kitaev chain with a T-junction}

Consider now a modified model where one of the chain ends is coupled
to an intermediate site forming a T-junction (sometimes also referred
to as a Y-junction), see Fig.~\ref{fig:modelT}. Such a model can be
easily formulated in the fermion language by adding another coupling
term to the Hamiltonian (\ref{hkit})
\begin{eqnarray}
\label{ht}
&&
\widehat{H} = -J\!\sum_{n=1}^{L-1} \!
\left(\hat c^\dagger_n \!-\! \hat c_n\right)\!
\left(\hat c^\dagger_{n+1} \!+\! \hat c_{n+1} \right)
\\
&&
\nonumber\\
&&
\qquad\qquad\quad
-J\left(\hat c^\dagger_L \!-\! \hat c_L\right)\!
\left(\hat c^\dagger_{N_0} \!+\! \hat c_{N_0} \right)
\nonumber\\
&&
\nonumber\\
&&
\qquad\qquad\qquad\qquad
-\sum_{n=1}^{L} h_n \left(\hat c^\dagger_n \!-\! \hat c_n\right)\!
\left(\hat c^\dagger_{n} \!+\! \hat c_{n} \right),
\nonumber
\end{eqnarray}
where
\[
1<N_0<L.
\]
A similar modification is also possible for the Ising Hamiltonian
(\ref{his}). However, such model cannot be mapped onto a quadratic
fermionic Hamiltonian similar to (\ref{ht}) due to non-cancellation of
the Jordan-Wigner strings at the junction point.

T-junctions play an important role in the literature on Majorana-based
quantum computation \cite{Fisher,Tewari,Akhmerov,Alicea}, where they
are the key elements of connected networks of quantum wires that are
envisioned to allow for braiding operations. A discussion of braiding
as well as any other time-dependent processes involving Majorana
fermions is beyond the scope of this paper.

\begin{figure}[t]
\centerline{\includegraphics[width=0.95\columnwidth]{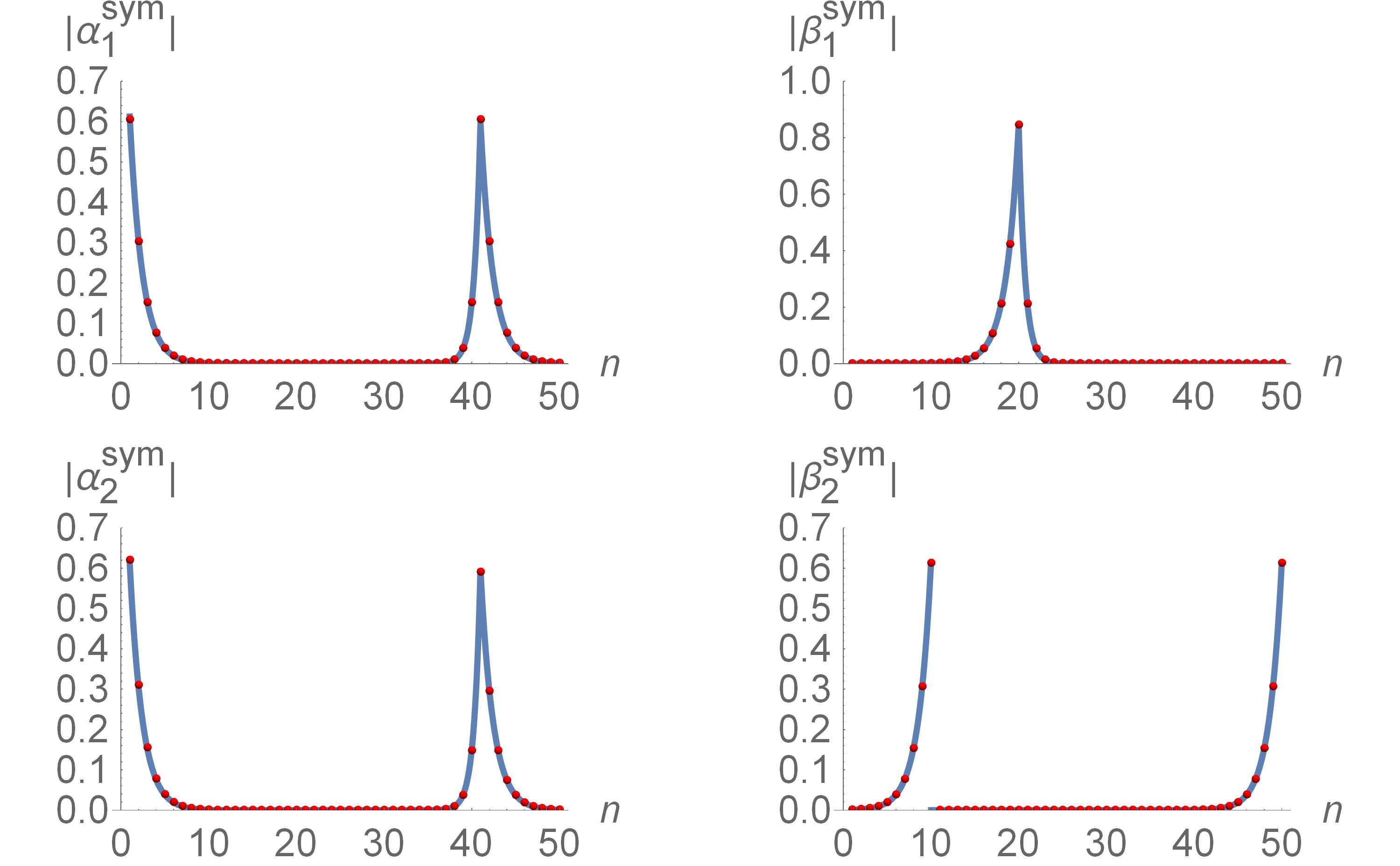}}
\caption{Majorana amplitudes of the two nearly zero-energy eigenstates
  of the modified Kitaev chain (\ref{ht}) with a T-junction in the
  piece-wise uniform applied field (\ref{hn}) with
  ${\lambda_1\!=\!\lambda_3\!=\!1/2}$, ${\lambda_2\!=\!4}$,
  ${N_1\!=\!M\!=\!20}$, ${N_2\!=\!10}$, and ${N_0\!=\!11}$. The red
  dots represent the result of the exact numerical
  diagonalization. The solid curves are presented for comparison. The
  amplitudes $\alpha_{1,2}^{\rm sym}$ shown on the left two panels are
  compared with Eq.~(\ref{asym}). The amplitude $\beta_1^{\rm sym}$
  appears to be well described by Eq.~(\ref{a2r}). The amplitude
  $\beta_2^{\rm sym}$ of the Majorana zero mode localized at the
  junction point is given by Eq.~(\ref{bt}).}
\label{fig:ab_T_20_10}
\end{figure}

The Hamiltonian (\ref{ht}) is quadratic and can be diagonalized by the
same exact method as the model (\ref{hkit}). Due to the more
complicated topology and larger number of parameters, the model
(\ref{ht}) exhibits many more parameter regimes than the chain
(\ref{hkit}). A comprehensive discussion of all of these regimes will
be presented elsewhere. In this paper, I focus on a single parameter
regime, where the nearly zero-energy eigenstates are described by the
amplitudes $\alpha_n$ that are delocalized between two interface
points similarly to Eq.~(\ref{asym}). In particular, I consider the
regime, where the junction site $N_0$ is approximately in the middle
of its ``topological'' region, which is twice as long as the second
``topological'' region. Choosing the equal parameters
${\lambda_1\!=\!\lambda_3\!<\!1}$ describing the ``topological''
regions, I achieve a configuration that is somewhat analogous to the
symmetric chain.

The results of the exact numerical diagonalization of the model
(\ref{ht}) in the chosen parameter regime (with ${N_1\!=\!20}$,
${M\!=\!20}$, ${N_2\!=\!10}$, and ${N_0\!=\!11}$) are presented in
Fig.~\ref{fig:ab_T_20_10} by the red dots and illustrated in
Fig.~\ref{fig:ab_T_3D}. The solution is characterized by the
amplitudes $\alpha^{\rm sym}_{1,2}$ that are delocalized between the
edge point of the chain and one of the borders of the ``trivial''
region. The amplitude $\alpha^{\rm sym}_{2}$ appears to be in perfect
agreement with Eq.~(\ref{asym}), while the amplitude $\alpha^{\rm
  sym}_{1}$ shows a barely perceptible deviation (see the solid curves
in the two left panels in Fig.~\ref{fig:ab_T_20_10}).

Now the amplitudes $\beta^{\rm sym}_{1,2}$ are no longer
delocalized. The amplitude $\beta^{\rm sym}_{1}$ is localized at the
edge of the ``trivial'' region and is perfectly described by
Eq.~(\ref{a2r}), see the top right panel in Fig.~\ref{fig:ab_T_20_10}.
The amplitude $\beta^{\rm sym}_{2}$ is localized at the T-junction
point, see the bottom right panel in Fig.~\ref{fig:ab_T_20_10}, and
can be described analytically by
\begin{equation}
\label{bt}
\beta^{\rm sym}_{2n}=\frac{(-1)^n}{\sqrt{2}}
\begin{cases}
s_1\lambda_1^{N_0\!-\!1\!-\!n}
,
& \!\!\!1\!\leqslant\! n \!\leqslant\! N_0\!-\!1, \cr
{\cal O}(\lambda^{N}), & \!\!\!N_0\!\leqslant\! n \!\leqslant\! N_1\!+\!M,\cr
-s_1\lambda_1^{L\!-\!n}
, 
& \!\!\!N_1\!+\!M\!+\!1\!\leqslant\! n \!\leqslant\! L.
\end{cases}
\end{equation}
The asymmetry of this amplitude -- $\beta^{\rm sym}_{2}$ is spread
over only two out of the three branches of the junction -- is related
to the chirality of the junction \cite{Tewari} described by the
explicitly asymmetric Hamiltonian (\ref{ht}), as well as to the time
reversal symmetry \cite{Ardonne} of the Hamiltonian (\ref{ht}).

 

\section{Discussion}

In this paper I presented the exact analytic solution of the
nonuniform Ising-Kitaev chain (\ref{h0}) with open boundary
conditions. The motivation for this work was two-fold: (i) I was
motivated by the proposal \cite{Mooij,Nori} of experimental
realization of zero-energy Majorana states in an artificial spin chain
engineered using Josephson qubits; such a system would be discreet
and, given current technological limitations, contain not too many
qubits; (ii) I wanted to reach a better understanding of the the
effective coupling between the Majorana zero modes in networked
systems used as paradigmatic examples of possible applications to
quantum computing \cite{Alicea}, in particular, the common
``nearest-neighbor'' form of the effective low-energy Hamiltonian
\cite{Shnirman,Nori15,Wang}. The model solved in this paper represents
the first step in reaching these goals.  In the case, where the two
``topological'' regions of the chain are separated by a ``trivial''
region, the exact analytic form of the eigenvalues and eigenvectors
can be found. Moreover, the resulting low-energy theory does not
support the ``nearest-neighbor'' approach \cite{Shnirman,Nori15,Wang}.

Indeed, in the generic parameter regime with the ``strong barrier'',
the single-fermion sector of the effective low-energy theory contains
two single-fermion states, each confined to its own ``topological''
region as if these regions were disconnected. Similarly to the
original Kitaev model \cite{Kitaev}, each of these states can be
interpreted in terms of two Majorana states localized at the edges of
the ``topological'' regions, as expected
\cite{Alicea,Shnirman,Nori15,Wang}. In the basis of the localized
Majorana fermions, the effective Hamiltonian has the block-diagonal
form (\ref{he1}). Here the two pairs of Majorana fermions $\gamma_1$,
$\gamma_2$ and $\gamma_3$, $\gamma_4$ form the two single-fermion
eigenstates. The orthogonality of the single-fermion eigenstates leads
to the absence of any coupling between $\gamma_2$ and $\gamma_3$, that
is typically included in the ``nearest-neighbor'' approach
\cite{Shnirman,Nori15,Wang}.

In contrast, in the specially fine-tuned case of the symmetric chain
the single-fermion eigenstates are equally spread between the two
``topological'' regions. Consequently, the corresponding Majorana
fermions are localized not at one, but at two separate interface
points. While the resulting low-energy Hamiltonian can of course be
represented in the above basis of localized Majorana states, the
latter are no longer related to the exact eigenstates of the model.
As a result, the Hamiltonian (\ref{esym}) exhibits additional
couplings between the two pairs $\gamma_2$, $\gamma_3$ and $\gamma_1$,
$\gamma_4$. These couplings are identical despite the large difference
in the separation between $\gamma_2$, $\gamma_3$ and $\gamma_1$,
$\gamma_4$. Again, this contradicts the ``nearest-neighbor'' approach
\cite{Shnirman,Nori15,Wang}, where the coupling between $\gamma_1$,
$\gamma_4$ is not included due to the larger separation (as compared
to other pairs of Majorana fermions).

Now, in the case of the weak barrier the amplitudes (\ref{a3r}) and
(\ref{a4r}) also exhibit the ``delocalization'' between two interface
points (although to a significantly lesser degree). Here, the dominant
terms in the effective Hamiltonian (\ref{he3}) are the couplings
between $\gamma_2$, $\gamma_3$ and $\gamma_1$, $\gamma_4$, while the
couplings $\gamma_1$, $\gamma_2$ and $\gamma_3$, $\gamma_4$ appear
with the typically small coefficient $\kappa$ (due to the weak
``delocalization''). As a result, the Hamiltonian (\ref{he3}) is also
incompatible with the ``nearest-neighbor'' form
\cite{Shnirman,Nori15,Wang}.

Finally, the ``delocalization'' of the Majorana zero modes between two
separated interface points is not an artifact \cite{Stern07} of the
model (\ref{hkit}). In particular, the modified model (\ref{ht}) also
exhibits this ``delocalization'' (again, requiring some fine-tuning).

The results of this paper are relevant for experimentalists designing
small systems hosting multiple zero-energy Majorana states
\cite{Mooij,Nori,Oreg13,Nori16}. In particular, in systems involving
relatively few Josephson qubits with conservative parameter values the
spreading of the Majorana zero modes over a few qubits and their
``delocalization'' between two well separated points are generic
effects \cite{Stern07} that need to be taken into account while
interpreting the experimental data and especially when planning any
kind of manipulation of the Majorana bound states by some external
bias. The ``delocalization'' of the Majorana zero modes and the
corresponding separation-independent effective coupling between some
pairs of the localized Majorana fermions could be observed in
experimental nanowire samples in the presence of additional symmetries
incorporated into the system design.

\begin{acknowledgments}

The author would like to thank Alexander Shnirman and Victor Gurarie
for numerous helpful discussions. I acknowledge support by the EU
Network Grant InterNoM and by Deutsche Forschungsgemeinschaft through
Grants No. SCHO 287/7-1, No. SH 81/2-1, and Open Access Publishing
Fund of Karlsruhe Institute of Technology, and the MEPhI Academic
Excellence Project (Contract No. 02.a03.21.0005).

\end{acknowledgments}


\begin{thebibliography}{53}
\expandafter\ifx\csname natexlab\endcsname\relax\def\natexlab#1{#1}\fi
\expandafter\ifx\csname bibnamefont\endcsname\relax
  \def\bibnamefont#1{#1}\fi
\expandafter\ifx\csname bibfnamefont\endcsname\relax
  \def\bibfnamefont#1{#1}\fi
\expandafter\ifx\csname citenamefont\endcsname\relax
  \def\citenamefont#1{#1}\fi
\expandafter\ifx\csname url\endcsname\relax
  \def\url#1{\texttt{#1}}\fi
\expandafter\ifx\csname urlprefix\endcsname\relax\def\urlprefix{URL }\fi
\providecommand{\bibinfo}[2]{#2}
\providecommand{\eprint}[2][]{\url{#2}}

\bibitem[{\citenamefont{Majorana}(1937)}]{Majorana}
\bibinfo{author}{\bibfnamefont{E.}~\bibnamefont{Majorana}},
  \bibinfo{journal}{Nuovo Cimento} \textbf{\bibinfo{volume}{14}},
  \bibinfo{pages}{171} (\bibinfo{year}{1937}).

\bibitem[{\citenamefont{Wilczek}(2008)}]{Wilczek}
\bibinfo{author}{\bibfnamefont{F.}~\bibnamefont{Wilczek}},
  \bibinfo{journal}{Nature Physics} \textbf{\bibinfo{volume}{5}},
  \bibinfo{pages}{614} (\bibinfo{year}{2008}).

\bibitem[{\citenamefont{Avignone et~al.}(2008)\citenamefont{Avignone, Elliott,
  and Engel}}]{Avignone}
\bibinfo{author}{\bibfnamefont{F.~T.} \bibnamefont{Avignone}},
  \bibinfo{author}{\bibfnamefont{S.~R.} \bibnamefont{Elliott}},
  \bibnamefont{and} \bibinfo{author}{\bibfnamefont{J.}~\bibnamefont{Engel}},
  \bibinfo{journal}{Rev. Mod. Phys.} \textbf{\bibinfo{volume}{80}},
  \bibinfo{pages}{481} (\bibinfo{year}{2008}).

\bibitem[{\citenamefont{Young et~al.}(2012)\citenamefont{Young, Zaheer, Teo,
  Kane, Mele, and Rappe}}]{Young}
\bibinfo{author}{\bibfnamefont{S.~M.} \bibnamefont{Young}},
  \bibinfo{author}{\bibfnamefont{S.}~\bibnamefont{Zaheer}},
  \bibinfo{author}{\bibfnamefont{J.~C.~Y.} \bibnamefont{Teo}},
  \bibinfo{author}{\bibfnamefont{C.~L.} \bibnamefont{Kane}},
  \bibinfo{author}{\bibfnamefont{E.~J.} \bibnamefont{Mele}}, \bibnamefont{and}
  \bibinfo{author}{\bibfnamefont{A.~M.} \bibnamefont{Rappe}},
  \bibinfo{journal}{Phys. Rev. Lett.} \textbf{\bibinfo{volume}{108}},
  \bibinfo{pages}{140405} (\bibinfo{year}{2012}).

\bibitem[{\citenamefont{Huang et~al.}(2015)\citenamefont{Huang, Xu, Belopolski,
  Lee, Chang, Wang, Alidoust, Bian, Neupane, Zhang et~al.}}]{Hasan}
\bibinfo{author}{\bibfnamefont{S.-M.} \bibnamefont{Huang}},
  \bibinfo{author}{\bibfnamefont{S.-Y.} \bibnamefont{Xu}},
  \bibinfo{author}{\bibfnamefont{I.}~\bibnamefont{Belopolski}},
  \bibinfo{author}{\bibfnamefont{C.-C.} \bibnamefont{Lee}},
  \bibinfo{author}{\bibfnamefont{G.}~\bibnamefont{Chang}},
  \bibinfo{author}{\bibfnamefont{B.}~\bibnamefont{Wang}},
  \bibinfo{author}{\bibfnamefont{N.}~\bibnamefont{Alidoust}},
  \bibinfo{author}{\bibfnamefont{G.}~\bibnamefont{Bian}},
  \bibinfo{author}{\bibfnamefont{M.}~\bibnamefont{Neupane}},
  \bibinfo{author}{\bibfnamefont{C.}~\bibnamefont{Zhang}},
  \bibnamefont{et~al.}, \bibinfo{journal}{Nature Comms.}
  \textbf{\bibinfo{volume}{6}}, \bibinfo{pages}{7373} (\bibinfo{year}{2015}).

\bibitem[{\citenamefont{Felser and Yan}(2016)}]{Yan}
\bibinfo{author}{\bibfnamefont{C.}~\bibnamefont{Felser}} \bibnamefont{and}
  \bibinfo{author}{\bibfnamefont{B.}~\bibnamefont{Yan}},
  \bibinfo{journal}{Nature Mat.} \textbf{\bibinfo{volume}{15}},
  \bibinfo{pages}{1149} (\bibinfo{year}{2016}).

\bibitem[{\citenamefont{Balents}(2011)}]{Balents}
\bibinfo{author}{\bibfnamefont{L.}~\bibnamefont{Balents}},
  \bibinfo{journal}{Physics} \textbf{\bibinfo{volume}{4}}, \bibinfo{pages}{36}
  (\bibinfo{year}{2011}).

\bibitem[{\citenamefont{Fang et~al.}(2012)\citenamefont{Fang, Gilbert, Dai, and
  Bernevig}}]{Bernevig}
\bibinfo{author}{\bibfnamefont{C.}~\bibnamefont{Fang}},
  \bibinfo{author}{\bibfnamefont{M.~J.} \bibnamefont{Gilbert}},
  \bibinfo{author}{\bibfnamefont{X.}~\bibnamefont{Dai}}, \bibnamefont{and}
  \bibinfo{author}{\bibfnamefont{B.~A.} \bibnamefont{Bernevig}},
  \bibinfo{journal}{Phys. Rev. Lett.} \textbf{\bibinfo{volume}{108}},
  \bibinfo{pages}{266802} (\bibinfo{year}{2012}).

\bibitem[{\citenamefont{Adler}(1969)}]{Adler}
\bibinfo{author}{\bibfnamefont{S.~L.} \bibnamefont{Adler}},
  \bibinfo{journal}{Phys. Rev.} \textbf{\bibinfo{volume}{177}},
  \bibinfo{pages}{2426} (\bibinfo{year}{1969}).

\bibitem[{\citenamefont{Bell and Jackiw}(1969)}]{Bell}
\bibinfo{author}{\bibfnamefont{J.~S.} \bibnamefont{Bell}} \bibnamefont{and}
  \bibinfo{author}{\bibfnamefont{R.}~\bibnamefont{Jackiw}},
  \bibinfo{journal}{Il Nuovo Cimento A} \textbf{\bibinfo{volume}{60}},
  \bibinfo{pages}{47} (\bibinfo{year}{1969}).

\bibitem[{\citenamefont{Nielsen and Ninomiya}(1983)}]{Nielsen}
\bibinfo{author}{\bibfnamefont{H.}~\bibnamefont{Nielsen}} \bibnamefont{and}
  \bibinfo{author}{\bibfnamefont{M.}~\bibnamefont{Ninomiya}},
  \bibinfo{journal}{Physics Letters B} \textbf{\bibinfo{volume}{130}},
  \bibinfo{pages}{389} (\bibinfo{year}{1983}).

\bibitem[{\citenamefont{Zhang et~al.}(2016)\citenamefont{Zhang, Xu, Belopolski,
  Yuan, Lin, Tong, Bian, Alidoust, Lee, Huang et~al.}}]{Zhang}
\bibinfo{author}{\bibfnamefont{C.-L.} \bibnamefont{Zhang}},
  \bibinfo{author}{\bibfnamefont{S.-Y.} \bibnamefont{Xu}},
  \bibinfo{author}{\bibfnamefont{I.}~\bibnamefont{Belopolski}},
  \bibinfo{author}{\bibfnamefont{Z.}~\bibnamefont{Yuan}},
  \bibinfo{author}{\bibfnamefont{Z.}~\bibnamefont{Lin}},
  \bibinfo{author}{\bibfnamefont{B.}~\bibnamefont{Tong}},
  \bibinfo{author}{\bibfnamefont{G.}~\bibnamefont{Bian}},
  \bibinfo{author}{\bibfnamefont{N.}~\bibnamefont{Alidoust}},
  \bibinfo{author}{\bibfnamefont{C.-C.} \bibnamefont{Lee}},
  \bibinfo{author}{\bibfnamefont{S.-M.} \bibnamefont{Huang}},
  \bibnamefont{et~al.}, \bibinfo{journal}{Nature Comm.}
  \textbf{\bibinfo{volume}{7}}, \bibinfo{pages}{10735} (\bibinfo{year}{2016}).

\bibitem[{\citenamefont{Mourik et~al.}(2012)\citenamefont{Mourik, Zuo, Frolov,
  Plissard, Bakkers, and Kouwenhoven}}]{Mourik}
\bibinfo{author}{\bibfnamefont{V.}~\bibnamefont{Mourik}},
  \bibinfo{author}{\bibfnamefont{K.}~\bibnamefont{Zuo}},
  \bibinfo{author}{\bibfnamefont{S.~M.} \bibnamefont{Frolov}},
  \bibinfo{author}{\bibfnamefont{S.~R.} \bibnamefont{Plissard}},
  \bibinfo{author}{\bibfnamefont{E.~P. A.~M.} \bibnamefont{Bakkers}},
  \bibnamefont{and} \bibinfo{author}{\bibfnamefont{L.~P.}
  \bibnamefont{Kouwenhoven}}, \bibinfo{journal}{Science}
  \textbf{\bibinfo{volume}{336}}, \bibinfo{pages}{1003} (\bibinfo{year}{2012}).

\bibitem[{\citenamefont{Rokhinson et~al.}(2012)\citenamefont{Rokhinson, Liu,
  and Furdyna}}]{Rokhinson}
\bibinfo{author}{\bibfnamefont{L.~P.} \bibnamefont{Rokhinson}},
  \bibinfo{author}{\bibfnamefont{X.}~\bibnamefont{Liu}}, \bibnamefont{and}
  \bibinfo{author}{\bibfnamefont{J.~K.} \bibnamefont{Furdyna}},
  \bibinfo{journal}{Nature Physics} \textbf{\bibinfo{volume}{8}},
  \bibinfo{pages}{795} (\bibinfo{year}{2012}).

\bibitem[{\citenamefont{Das et~al.}(2012)\citenamefont{Das, Ronen, Most, Oreg,
  Heiblum, and Shtrikman}}]{Heiblum}
\bibinfo{author}{\bibfnamefont{A.}~\bibnamefont{Das}},
  \bibinfo{author}{\bibfnamefont{Y.}~\bibnamefont{Ronen}},
  \bibinfo{author}{\bibfnamefont{Y.}~\bibnamefont{Most}},
  \bibinfo{author}{\bibfnamefont{Y.}~\bibnamefont{Oreg}},
  \bibinfo{author}{\bibfnamefont{M.}~\bibnamefont{Heiblum}}, \bibnamefont{and}
  \bibinfo{author}{\bibfnamefont{H.}~\bibnamefont{Shtrikman}},
  \bibinfo{journal}{Nature Physics} \textbf{\bibinfo{volume}{8}},
  \bibinfo{pages}{887} (\bibinfo{year}{2012}).

\bibitem[{\citenamefont{Deng et~al.}(2012)\citenamefont{Deng, Yu, Huang,
  Larsson, Caroff, and Xu}}]{Deng}
\bibinfo{author}{\bibfnamefont{M.~T.} \bibnamefont{Deng}},
  \bibinfo{author}{\bibfnamefont{C.~L.} \bibnamefont{Yu}},
  \bibinfo{author}{\bibfnamefont{G.~Y.} \bibnamefont{Huang}},
  \bibinfo{author}{\bibfnamefont{M.}~\bibnamefont{Larsson}},
  \bibinfo{author}{\bibfnamefont{P.}~\bibnamefont{Caroff}}, \bibnamefont{and}
  \bibinfo{author}{\bibfnamefont{H.~Q.} \bibnamefont{Xu}},
  \bibinfo{journal}{Nano Letters} \textbf{\bibinfo{volume}{12}},
  \bibinfo{pages}{6414} (\bibinfo{year}{2012}).

\bibitem[{\citenamefont{Churchill et~al.}(2013)\citenamefont{Churchill, Fatemi,
  Grove-Rasmussen, Deng, Caroff, Xu, and Marcus}}]{Marcus}
\bibinfo{author}{\bibfnamefont{H.~O.~H.} \bibnamefont{Churchill}},
  \bibinfo{author}{\bibfnamefont{V.}~\bibnamefont{Fatemi}},
  \bibinfo{author}{\bibfnamefont{K.}~\bibnamefont{Grove-Rasmussen}},
  \bibinfo{author}{\bibfnamefont{M.~T.} \bibnamefont{Deng}},
  \bibinfo{author}{\bibfnamefont{P.}~\bibnamefont{Caroff}},
  \bibinfo{author}{\bibfnamefont{H.~Q.} \bibnamefont{Xu}}, \bibnamefont{and}
  \bibinfo{author}{\bibfnamefont{C.~M.} \bibnamefont{Marcus}},
  \bibinfo{journal}{Phys. Rev. B} \textbf{\bibinfo{volume}{87}},
  \bibinfo{pages}{241401} (\bibinfo{year}{2013}).

\bibitem[{\citenamefont{Lee et~al.}(2014)\citenamefont{Lee, Jiang, Houzet,
  Aguado, Lieber, and {De Franceschi}}}]{DeFranceschi}
\bibinfo{author}{\bibfnamefont{E.~J.} \bibnamefont{Lee}},
  \bibinfo{author}{\bibfnamefont{X.}~\bibnamefont{Jiang}},
  \bibinfo{author}{\bibfnamefont{M.}~\bibnamefont{Houzet}},
  \bibinfo{author}{\bibfnamefont{R.}~\bibnamefont{Aguado}},
  \bibinfo{author}{\bibfnamefont{C.~M.} \bibnamefont{Lieber}},
  \bibnamefont{and} \bibinfo{author}{\bibfnamefont{S.}~\bibnamefont{{De
  Franceschi}}}, \bibinfo{journal}{Nature Nanotechnology}
  \textbf{\bibinfo{volume}{9}}, \bibinfo{pages}{79} (\bibinfo{year}{2014}).

\bibitem[{\citenamefont{Albrecht1 et~al.}(2016)\citenamefont{Albrecht1,
  Higginbotham, Madsen, Kuemmeth, Jespersen, Nygard, Krogstrup1, and
  Marcus}}]{Albrecht}
\bibinfo{author}{\bibfnamefont{S.~M.} \bibnamefont{Albrecht1}},
  \bibinfo{author}{\bibfnamefont{A.~P.} \bibnamefont{Higginbotham}},
  \bibinfo{author}{\bibfnamefont{M.}~\bibnamefont{Madsen}},
  \bibinfo{author}{\bibfnamefont{F.}~\bibnamefont{Kuemmeth}},
  \bibinfo{author}{\bibfnamefont{T.~S.} \bibnamefont{Jespersen}},
  \bibinfo{author}{\bibfnamefont{J.}~\bibnamefont{Nygard}},
  \bibinfo{author}{\bibfnamefont{P.}~\bibnamefont{Krogstrup1}},
  \bibnamefont{and} \bibinfo{author}{\bibfnamefont{C.~M.}
  \bibnamefont{Marcus}}, \bibinfo{journal}{Nature}
  \textbf{\bibinfo{volume}{531}}, \bibinfo{pages}{206} (\bibinfo{year}{2016}).

\bibitem[{\citenamefont{Deng et~al.}(2016)\citenamefont{Deng, Vaitiekenas,
  Hansen, Danon, Leijnse, Flensberg, Nygard, Krogstrup, and Marcus}}]{Marcus16}
\bibinfo{author}{\bibfnamefont{M.~T.} \bibnamefont{Deng}},
  \bibinfo{author}{\bibfnamefont{S.}~\bibnamefont{Vaitiekenas}},
  \bibinfo{author}{\bibfnamefont{E.~B.} \bibnamefont{Hansen}},
  \bibinfo{author}{\bibfnamefont{J.}~\bibnamefont{Danon}},
  \bibinfo{author}{\bibfnamefont{M.}~\bibnamefont{Leijnse}},
  \bibinfo{author}{\bibfnamefont{K.}~\bibnamefont{Flensberg}},
  \bibinfo{author}{\bibfnamefont{J.}~\bibnamefont{Nygard}},
  \bibinfo{author}{\bibfnamefont{P.}~\bibnamefont{Krogstrup}},
  \bibnamefont{and} \bibinfo{author}{\bibfnamefont{C.~M.}
  \bibnamefont{Marcus}}, \bibinfo{journal}{Science}
  \textbf{\bibinfo{volume}{354}}, \bibinfo{pages}{1557} (\bibinfo{year}{2016}).

\bibitem[{\citenamefont{Kitaev}(2001)}]{Kitaev}
\bibinfo{author}{\bibfnamefont{A.~Y.} \bibnamefont{Kitaev}},
  \bibinfo{journal}{Physics-Uspekhi} \textbf{\bibinfo{volume}{44}},
  \bibinfo{pages}{131} (\bibinfo{year}{2001}).

\bibitem[{\citenamefont{Fu and Kane}(2008)}]{Kane}
\bibinfo{author}{\bibfnamefont{L.}~\bibnamefont{Fu}} \bibnamefont{and}
  \bibinfo{author}{\bibfnamefont{C.~L.} \bibnamefont{Kane}},
  \bibinfo{journal}{Phys. Rev. Lett.} \textbf{\bibinfo{volume}{100}},
  \bibinfo{pages}{096407} (\bibinfo{year}{2008}).

\bibitem[{\citenamefont{Lutchyn et~al.}(2010)\citenamefont{Lutchyn, Sau, and
  Das~Sarma}}]{Lutchyn}
\bibinfo{author}{\bibfnamefont{R.~M.} \bibnamefont{Lutchyn}},
  \bibinfo{author}{\bibfnamefont{J.~D.} \bibnamefont{Sau}}, \bibnamefont{and}
  \bibinfo{author}{\bibfnamefont{S.}~\bibnamefont{Das~Sarma}},
  \bibinfo{journal}{Phys. Rev. Lett.} \textbf{\bibinfo{volume}{105}},
  \bibinfo{pages}{077001} (\bibinfo{year}{2010}).

\bibitem[{\citenamefont{Oreg et~al.}(2010)\citenamefont{Oreg, Refael, and von
  Oppen}}]{Oreg}
\bibinfo{author}{\bibfnamefont{Y.}~\bibnamefont{Oreg}},
  \bibinfo{author}{\bibfnamefont{G.}~\bibnamefont{Refael}}, \bibnamefont{and}
  \bibinfo{author}{\bibfnamefont{F.}~\bibnamefont{von Oppen}},
  \bibinfo{journal}{Phys. Rev. Lett.} \textbf{\bibinfo{volume}{105}},
  \bibinfo{pages}{177002} (\bibinfo{year}{2010}).

\bibitem[{\citenamefont{Barber}(1983)}]{Barber}
\bibinfo{author}{\bibfnamefont{M.~N.} \bibnamefont{Barber}}
  (\bibinfo{publisher}{Academic Press, London}, \bibinfo{year}{1983}),
  vol.~\bibinfo{volume}{8} of \emph{\bibinfo{series}{Phase transitions and
  critical phenomena}}.

\bibitem[{\citenamefont{Beenakker}(2013)}]{Beenakker}
\bibinfo{author}{\bibfnamefont{C.}~\bibnamefont{Beenakker}},
  \bibinfo{journal}{Annual Review of Condensed Matter Physics}
  \textbf{\bibinfo{volume}{4}}, \bibinfo{pages}{113} (\bibinfo{year}{2013}).

\bibitem[{\citenamefont{Alicea}(2012)}]{Alicea}
\bibinfo{author}{\bibfnamefont{J.}~\bibnamefont{Alicea}},
  \bibinfo{journal}{Reports on Progress in Physics}
  \textbf{\bibinfo{volume}{75}}, \bibinfo{pages}{076501}
  (\bibinfo{year}{2012}).

\bibitem[{\citenamefont{Kitaev}(2010)}]{Kitaev08}
\bibinfo{author}{\bibfnamefont{A.~Y.} \bibnamefont{Kitaev}}, in
  \emph{\bibinfo{booktitle}{Exact Methods in Low-dimensional Statistical
  Physics and Quantum Computing}}, edited by
  \bibinfo{editor}{\bibfnamefont{J.}~\bibnamefont{Jacobsen}},
  \bibinfo{editor}{\bibfnamefont{S.}~\bibnamefont{Ouvry}},
  \bibinfo{editor}{\bibfnamefont{V.}~\bibnamefont{Pasquier}},
  \bibinfo{editor}{\bibfnamefont{D.}~\bibnamefont{Serban}}, \bibnamefont{and}
  \bibinfo{editor}{\bibfnamefont{L.}~\bibnamefont{Cugliandolo}}
  (\bibinfo{publisher}{Oxford Univeristy Press}, \bibinfo{year}{2010}),
  vol.~\bibinfo{volume}{89} of \emph{\bibinfo{series}{Lecture Notes of the Les
  Houches Summer School}}, \bibinfo{note}{arXiv:0904.2771}.

\bibitem[{\citenamefont{Nayak et~al.}(2008)\citenamefont{Nayak, Simon, Stern,
  Freedman, and Das~Sarma}}]{Nayak}
\bibinfo{author}{\bibfnamefont{C.}~\bibnamefont{Nayak}},
  \bibinfo{author}{\bibfnamefont{S.~H.} \bibnamefont{Simon}},
  \bibinfo{author}{\bibfnamefont{A.}~\bibnamefont{Stern}},
  \bibinfo{author}{\bibfnamefont{M.}~\bibnamefont{Freedman}}, \bibnamefont{and}
  \bibinfo{author}{\bibfnamefont{S.}~\bibnamefont{Das~Sarma}},
  \bibinfo{journal}{Rev. Mod. Phys.} \textbf{\bibinfo{volume}{80}},
  \bibinfo{pages}{1083} (\bibinfo{year}{2008}).

\bibitem[{\citenamefont{Zhang and Nori}(2015)}]{Nori15}
\bibinfo{author}{\bibfnamefont{P.}~\bibnamefont{Zhang}} \bibnamefont{and}
  \bibinfo{author}{\bibfnamefont{F.}~\bibnamefont{Nori}},
  \bibinfo{journal}{Phys. Rev. B} \textbf{\bibinfo{volume}{92}},
  \bibinfo{pages}{115303} (\bibinfo{year}{2015}).

\bibitem[{\citenamefont{Stern}(2008)}]{Stern}
\bibinfo{author}{\bibfnamefont{A.}~\bibnamefont{Stern}},
  \bibinfo{journal}{Annals of Physics} \textbf{\bibinfo{volume}{323}},
  \bibinfo{pages}{204} (\bibinfo{year}{2008}).

\bibitem[{\citenamefont{Halperin et~al.}(2012)\citenamefont{Halperin, Oreg,
  Stern, Refael, Alicea, and von Oppen}}]{Halperin}
\bibinfo{author}{\bibfnamefont{B.~I.} \bibnamefont{Halperin}},
  \bibinfo{author}{\bibfnamefont{Y.}~\bibnamefont{Oreg}},
  \bibinfo{author}{\bibfnamefont{A.}~\bibnamefont{Stern}},
  \bibinfo{author}{\bibfnamefont{G.}~\bibnamefont{Refael}},
  \bibinfo{author}{\bibfnamefont{J.}~\bibnamefont{Alicea}}, \bibnamefont{and}
  \bibinfo{author}{\bibfnamefont{F.}~\bibnamefont{von Oppen}},
  \bibinfo{journal}{Phys. Rev. B} \textbf{\bibinfo{volume}{85}},
  \bibinfo{pages}{144501} (\bibinfo{year}{2012}).

\bibitem[{\citenamefont{Levitov et~al.}(2001)\citenamefont{Levitov, Orlando,
  Majer, and Mooij}}]{Mooij}
\bibinfo{author}{\bibfnamefont{L.~S.} \bibnamefont{Levitov}},
  \bibinfo{author}{\bibfnamefont{T.~P.} \bibnamefont{Orlando}},
  \bibinfo{author}{\bibfnamefont{J.~B.} \bibnamefont{Majer}}, \bibnamefont{and}
  \bibinfo{author}{\bibfnamefont{J.~E.} \bibnamefont{Mooij}}
  (\bibinfo{year}{2001}), \bibinfo{note}{{c}ond-mat/0108266 (unpublished)}.

\bibitem[{\citenamefont{You et~al.}(2014)\citenamefont{You, Wang, Zhang, and
  Nori}}]{Nori}
\bibinfo{author}{\bibfnamefont{J.~Q.} \bibnamefont{You}},
  \bibinfo{author}{\bibfnamefont{Z.~D.} \bibnamefont{Wang}},
  \bibinfo{author}{\bibfnamefont{W.}~\bibnamefont{Zhang}}, \bibnamefont{and}
  \bibinfo{author}{\bibfnamefont{F.}~\bibnamefont{Nori}},
  \bibinfo{journal}{Scientific Reports} \textbf{\bibinfo{volume}{4}},
  \bibinfo{pages}{5535} (\bibinfo{year}{2014}).

\bibitem[{\citenamefont{Fulga et~al.}(2013)\citenamefont{Fulga, Haim, Akhmerov,
  and Oreg}}]{Oreg13}
\bibinfo{author}{\bibfnamefont{I.~C.} \bibnamefont{Fulga}},
  \bibinfo{author}{\bibfnamefont{A.}~\bibnamefont{Haim}},
  \bibinfo{author}{\bibfnamefont{A.~R.} \bibnamefont{Akhmerov}},
  \bibnamefont{and} \bibinfo{author}{\bibfnamefont{Y.}~\bibnamefont{Oreg}},
  \bibinfo{journal}{New Journal of Physics} \textbf{\bibinfo{volume}{15}},
  \bibinfo{pages}{045020} (\bibinfo{year}{2013}).

\bibitem[{\citenamefont{Zhang and Nori}(2016)}]{Nori16}
\bibinfo{author}{\bibfnamefont{P.}~\bibnamefont{Zhang}} \bibnamefont{and}
  \bibinfo{author}{\bibfnamefont{F.}~\bibnamefont{Nori}}, \bibinfo{journal}{New
  Journal of Physics} \textbf{\bibinfo{volume}{18}}, \bibinfo{pages}{043033}
  (\bibinfo{year}{2016}).

\bibitem[{\citenamefont{Ioffe et~al.}(2002)\citenamefont{Ioffe, Feigel'man,
  Ioselevich, Ivanov, Troyer, and Blatter}}]{Ioffe}
\bibinfo{author}{\bibfnamefont{L.~B.} \bibnamefont{Ioffe}},
  \bibinfo{author}{\bibfnamefont{M.~V.} \bibnamefont{Feigel'man}},
  \bibinfo{author}{\bibfnamefont{A.}~\bibnamefont{Ioselevich}},
  \bibinfo{author}{\bibfnamefont{D.}~\bibnamefont{Ivanov}},
  \bibinfo{author}{\bibfnamefont{M.}~\bibnamefont{Troyer}}, \bibnamefont{and}
  \bibinfo{author}{\bibfnamefont{G.}~\bibnamefont{Blatter}},
  \bibinfo{journal}{Nature} \textbf{\bibinfo{volume}{415}},
  \bibinfo{pages}{503} (\bibinfo{year}{2002}).

\bibitem[{\citenamefont{You and Nori}(2011)}]{Nori11}
\bibinfo{author}{\bibfnamefont{J.~Q.} \bibnamefont{You}} \bibnamefont{and}
  \bibinfo{author}{\bibfnamefont{F.}~\bibnamefont{Nori}},
  \bibinfo{journal}{Nature} \textbf{\bibinfo{volume}{474}},
  \bibinfo{pages}{589} (\bibinfo{year}{2011}).

\bibitem[{\citenamefont{Suzuki et~al.}(2013)\citenamefont{Suzuki, Inoue, and
  Chakrabarti}}]{Suzuki}
\bibinfo{author}{\bibfnamefont{S.}~\bibnamefont{Suzuki}},
  \bibinfo{author}{\bibfnamefont{J.}~\bibnamefont{Inoue}}, \bibnamefont{and}
  \bibinfo{author}{\bibfnamefont{B.~K.} \bibnamefont{Chakrabarti}},
  \emph{\bibinfo{title}{Quantum {I}sing phases and transitions in transverse
  {I}sing models}} (\bibinfo{publisher}{Springer}, \bibinfo{year}{2013}).

\bibitem[{\citenamefont{Gogolin et~al.}(1998)\citenamefont{Gogolin, Nersesyan,
  and Tsvelik}}]{Shura}
\bibinfo{author}{\bibfnamefont{A.~O.} \bibnamefont{Gogolin}},
  \bibinfo{author}{\bibfnamefont{A.~A.} \bibnamefont{Nersesyan}},
  \bibnamefont{and} \bibinfo{author}{\bibfnamefont{A.~M.}
  \bibnamefont{Tsvelik}}, \emph{\bibinfo{title}{Bosonization and Strongly
  Correlated Systems}} (\bibinfo{publisher}{Cambridge University Press},
  \bibinfo{year}{1998}).

\bibitem[{\citenamefont{Tsvelik}(2003)}]{Tsvelik}
\bibinfo{author}{\bibfnamefont{A.~M.} \bibnamefont{Tsvelik}},
  \emph{\bibinfo{title}{Quantum field theory in condensed matter physics}}
  (\bibinfo{publisher}{Cambridge University Press}, \bibinfo{year}{2003}).

\bibitem[{\citenamefont{Lieb et~al.}(1961)\citenamefont{Lieb, Schultz, and
  Mattis}}]{Lieb}
\bibinfo{author}{\bibfnamefont{E.}~\bibnamefont{Lieb}},
  \bibinfo{author}{\bibfnamefont{T.}~\bibnamefont{Schultz}}, \bibnamefont{and}
  \bibinfo{author}{\bibfnamefont{D.}~\bibnamefont{Mattis}},
  \bibinfo{journal}{Annals of Physics} \textbf{\bibinfo{volume}{16}},
  \bibinfo{pages}{407} (\bibinfo{year}{1961}).

\bibitem[{\citenamefont{Pfeuty}(1970)}]{Pfeuty}
\bibinfo{author}{\bibfnamefont{P.}~\bibnamefont{Pfeuty}},
  \bibinfo{journal}{Annals of Physics} \textbf{\bibinfo{volume}{57}},
  \bibinfo{pages}{79} (\bibinfo{year}{1970}).

\bibitem[{\citenamefont{Greiter et~al.}(2014)\citenamefont{Greiter, Schnells,
  and Thomale}}]{Greiter}
\bibinfo{author}{\bibfnamefont{M.}~\bibnamefont{Greiter}},
  \bibinfo{author}{\bibfnamefont{V.}~\bibnamefont{Schnells}}, \bibnamefont{and}
  \bibinfo{author}{\bibfnamefont{R.}~\bibnamefont{Thomale}},
  \bibinfo{journal}{Annals of Physics} \textbf{\bibinfo{volume}{351}},
  \bibinfo{pages}{1026} (\bibinfo{year}{2014}).

\bibitem[{\citenamefont{Jordan and Wigner}(1928)}]{JW}
\bibinfo{author}{\bibfnamefont{P.}~\bibnamefont{Jordan}} \bibnamefont{and}
  \bibinfo{author}{\bibfnamefont{E.}~\bibnamefont{Wigner}},
  \bibinfo{journal}{Zeitschrift f{\"u}r Physik} \textbf{\bibinfo{volume}{47}},
  \bibinfo{pages}{631} (\bibinfo{year}{1928}).

\bibitem[{\citenamefont{Bogolyubov}(1947)}]{Bog}
\bibinfo{author}{\bibfnamefont{N.~N.} \bibnamefont{Bogolyubov}},
  \bibinfo{journal}{Izv. Acad. Nauk USSR} \textbf{\bibinfo{volume}{11}},
  \bibinfo{pages}{77} (\bibinfo{year}{1947}), \bibinfo{note}{[J. Phys. (USSR)
  {\bf 11}, 23 (1947)]}.

\bibitem[{\citenamefont{Burnell et~al.}(2013)\citenamefont{Burnell, Shnirman,
  and Oreg}}]{Shnirman}
\bibinfo{author}{\bibfnamefont{F.~J.} \bibnamefont{Burnell}},
  \bibinfo{author}{\bibfnamefont{A.}~\bibnamefont{Shnirman}}, \bibnamefont{and}
  \bibinfo{author}{\bibfnamefont{Y.}~\bibnamefont{Oreg}},
  \bibinfo{journal}{Phys. Rev. B} \textbf{\bibinfo{volume}{88}},
  \bibinfo{pages}{224507} (\bibinfo{year}{2013}).

\bibitem[{\citenamefont{Wang et~al.}(2015)\citenamefont{Wang, Liang, Yao, and
  Hu}}]{Wang}
\bibinfo{author}{\bibfnamefont{Z.}~\bibnamefont{Wang}},
  \bibinfo{author}{\bibfnamefont{Q.-F.} \bibnamefont{Liang}},
  \bibinfo{author}{\bibfnamefont{D.-X.} \bibnamefont{Yao}}, \bibnamefont{and}
  \bibinfo{author}{\bibfnamefont{X.}~\bibnamefont{Hu}},
  \bibinfo{journal}{Scientific Reports} \textbf{\bibinfo{volume}{5}},
  \bibinfo{pages}{11686} (\bibinfo{year}{2015}).

\bibitem[{\citenamefont{Alicea et~al.}(2011)\citenamefont{Alicea, Oreg, Refael,
  von Oppen, and Fisher}}]{Fisher}
\bibinfo{author}{\bibfnamefont{J.}~\bibnamefont{Alicea}},
  \bibinfo{author}{\bibfnamefont{Y.}~\bibnamefont{Oreg}},
  \bibinfo{author}{\bibfnamefont{G.}~\bibnamefont{Refael}},
  \bibinfo{author}{\bibfnamefont{F.}~\bibnamefont{von Oppen}},
  \bibnamefont{and} \bibinfo{author}{\bibfnamefont{M.~P.~A.}
  \bibnamefont{Fisher}}, \bibinfo{journal}{Nature Physics}
  \textbf{\bibinfo{volume}{7}}, \bibinfo{pages}{412} (\bibinfo{year}{2011}).

\bibitem[{\citenamefont{Clarke et~al.}(2011)\citenamefont{Clarke, Sau, and
  Tewari}}]{Tewari}
\bibinfo{author}{\bibfnamefont{D.~J.} \bibnamefont{Clarke}},
  \bibinfo{author}{\bibfnamefont{J.~D.} \bibnamefont{Sau}}, \bibnamefont{and}
  \bibinfo{author}{\bibfnamefont{S.}~\bibnamefont{Tewari}},
  \bibinfo{journal}{Phys. Rev. B} \textbf{\bibinfo{volume}{84}},
  \bibinfo{pages}{035120} (\bibinfo{year}{2011}).

\bibitem[{\citenamefont{Hyart et~al.}(2013)\citenamefont{Hyart, van Heck,
  Fulga, Burrello, Akhmerov, and Beenakker}}]{Akhmerov}
\bibinfo{author}{\bibfnamefont{T.}~\bibnamefont{Hyart}},
  \bibinfo{author}{\bibfnamefont{B.}~\bibnamefont{van Heck}},
  \bibinfo{author}{\bibfnamefont{I.~C.} \bibnamefont{Fulga}},
  \bibinfo{author}{\bibfnamefont{M.}~\bibnamefont{Burrello}},
  \bibinfo{author}{\bibfnamefont{A.~R.} \bibnamefont{Akhmerov}},
  \bibnamefont{and} \bibinfo{author}{\bibfnamefont{C.~W.~J.}
  \bibnamefont{Beenakker}}, \bibinfo{journal}{Phys. Rev. B}
  \textbf{\bibinfo{volume}{88}}, \bibinfo{pages}{035121}
  (\bibinfo{year}{2013}).

\bibitem[{\citenamefont{Sp{\aa}nsl{\"a}tt and Ardonne}(2017)}]{Ardonne}
\bibinfo{author}{\bibfnamefont{C.}~\bibnamefont{Sp{\aa}nsl{\"a}tt}}
  \bibnamefont{and} \bibinfo{author}{\bibfnamefont{E.}~\bibnamefont{Ardonne}},
  \bibinfo{journal}{Journal of Physics: Condensed Matter}
  \textbf{\bibinfo{volume}{29}}, \bibinfo{pages}{105602}
  (\bibinfo{year}{2017}).

\bibitem[{\citenamefont{Grosfeld et~al.}(2007)\citenamefont{Grosfeld, Cooper,
  Stern, and Ilan}}]{Stern07}
\bibinfo{author}{\bibfnamefont{E.}~\bibnamefont{Grosfeld}},
  \bibinfo{author}{\bibfnamefont{N.~R.} \bibnamefont{Cooper}},
  \bibinfo{author}{\bibfnamefont{A.}~\bibnamefont{Stern}}, \bibnamefont{and}
  \bibinfo{author}{\bibfnamefont{R.}~\bibnamefont{Ilan}},
  \bibinfo{journal}{Phys. Rev. B} \textbf{\bibinfo{volume}{76}},
  \bibinfo{pages}{104516} (\bibinfo{year}{2007}).

\end{thebibliography}

\end{document}